\documentclass[12pt]{article}
\usepackage{color}
\usepackage{epsfig}
\usepackage{latexsym}
\usepackage{amsmath}
\usepackage{amssymb}
\usepackage{amsfonts}
\usepackage{pstricks}
\newcommand{\be}{\begin{equation}}
\newcommand{\ee}{\end{equation}}
\newcommand{\bea}{\begin{eqnarray}}
\newcommand{\eea}{\end{eqnarray}}

\newcommand{\ep}{\epsilon}

\newcommand{\nn}{\nonumber}

\begin{document}
\parindent=1.5pc

\begin{titlepage}

\bigskip
\begin{center}
{{\large\bf
Analytic Epsilon Expansions
of Master Integrals Corresponding to Massless Three-Loop Form Factors
and  Three-Loop  $g-2$ up to Four-Loop Transcendentality Weight
} \\
\vglue 5pt \vglue 1.0cm
{\large   R.N. Lee}\footnote{E-mail: R.N.Lee@inp.nsk.su}\\
\baselineskip=14pt \vspace{2mm} {\normalsize Budker Institute of Nuclear Physics and
 Novosibirsk State University, \\630090, Novosibirsk, Russia
}\\
\baselineskip=14pt \vspace{2mm} \baselineskip=14pt \vspace{2mm}
{\large   V.A. Smirnov}\footnote{E-mail: smirnov@theory.sinp.msu.ru}\\
\baselineskip=14pt \vspace{2mm} {\normalsize
Skobeltsyn Institute of Nuclear Physics of Moscow State University, \\
119992 Moscow, Russia
}\\
\baselineskip=14pt \vspace{2mm} \vglue 0.8cm {Abstract}}
\end{center}
\vglue 0.3cm {\rightskip=3pc
 \leftskip=3pc
\noindent We evaluate analytically higher terms of the $\ep$-expansion of the three-loop
master integrals corresponding to three-loop quark and gluon form factors and to the
three-loop master integrals contributing to the electron $g-2$ in QED
up to the transcendentality weight typical to four-loop calculations, i.e. eight
and seven, respectively.
The calculation is based on a combination of a method recently
suggested by one of the authors (R.L.) with other techniques: sector decomposition
implemented in {\tt FIESTA}, the method of Mellin--Barnes representation, and the PSLQ
algorithm.
\vglue 0.8cm}
\end{titlepage}

\section{Introduction}

One year ago a method of multiloop calculations \cite{Lee:2009dh} based on the use of
dimensional recurrence relations \cite{Tarasov1996} and analytic properties of Feynman
integrals as functions of the parameter of dimensional regularization, $d=4-2\ep$, was
suggested (the DRA method).  Then it was successfully applied \cite{Lee:2010cga} to the
analytic evaluation of the terms of order $\ep^0$ of the $\ep$-expansion of two most
complicated master integrals corresponding to the three-loop quark and gluon form factors.
These results provided the possibility to obtain completely analytic expressions for the form
factors. The power of the DRA method was further demonstrated in
Ref.~\cite{Lee:2010ug} where the three previously unknown terms of the expansion in $\ep$,
i.e. up to transcendentality weight ten, of the three-loop non-planar massless propagator
diagram were evaluated. Some additional details of the method were presented in
Ref.~\cite{Lee:2010we}.

The main point of Ref.~\cite{Lee:2010ug} was to emphasize that once
analytic expressions, in terms of well convergent series, are obtained within the DRA method
it is easy to evaluate extra terms of the expansion in $\ep$.
In the present paper we further exploit this nice feature of the method by evaluating,
up to the transcendentality weight typical to four-loop calculations,
higher terms of the $\ep$-expansion of the three-loop master integrals corresponding to
three-loop quark and gluon form factors and to the three-loop master integrals
contributing to the electron $g-2$ in QED.

In the first of these two problems, all the master integrals apart from three most complicated
ones were  evaluated in Refs.~\cite{3lff1,3lff2}. In fact, the word {\em evaluated} means here
the evaluation up to the order of $\ep$ which appears in the finite part of the form factors.
Mathematically, this means the evaluation up to transcendentality weight six. Then one of the
three most complicated master integrals (called $A_{9,1}$ in Refs.~\cite{3lff1,3lff2} and in the
present paper) and the pole parts of $A_{9,4}$ and $A_{9,2}$ (shown in Fig.~1 in the next
section) were evaluated analytically, while the $\ep^0$ parts of $A_{9,4}$ and $A_{9,2}$ were
evaluated numerically --- see Refs.~\cite{3lff3,3lff4}. These two missing ingredients were
calculated in Ref.~\cite{Lee:2010cga}. Motivated by future four-loop calculations, the authors
of Ref.~\cite{Gehrmann:2010ue} presented one more term of the $\ep$-expansion for all the
master integrals but $A_{9,1}$, $A_{9,2}$ and $A_{9,4}$. For $A_{9,1}$, this was done in
\cite{Lee:2010ug}. In the present paper we evaluate all the master integrals up to
transcendentality weight eight which is intrinsic to a four-loop evaluation of the form factors.
Explicitly, the constants present in results for highest powers of $\ep$ are linear combinations
of $\pi^8, \zeta_{3}^2\pi^2, \zeta_{5}\zeta_{3}$ and $\zeta_{-6,-2}$.

The second family of master integrals we are evaluating is connected with the anomalous
magnetic moment of the muon, i.e. $g - 2$ factor. The three-loop evaluation was performed in
Refs.~\cite{Laporta:1996mq,Laporta:1997zy}. The evaluation at the four-loop level, within a
pure numerical approach where each of the Feynman integrals involved is evaluated
numerically, without a reduction to master integrals, was already performed --- see
Ref.~\cite{Aoyama:2007dv} and references therein. Even the numerical evaluation in five
loops has been started --- see a very recent paper \cite{Aoyama:2010pk} and references
therein.

In Ref.~\cite{Laporta:2008zz} the status of partially analytic evaluation of the four-loop $g-2$
factor was presented. In this approach, an integration by parts (IBP) reduction \cite{IBP}
to master integrals is used.
However, it was pointed out that the analytic evaluation of the corresponding master integrals
was not possible for the moment. To evaluate the master integrals numerically with a high
precision it was supposed to use the method developed in
Refs.~\cite{Laporta:2001dd,Laporta:2000dc} and based on difference equations. Keeping in
mind that all the three-loop $g-2$ master integrals will inevitably appear in a four-loop
calculation because they are present in factorizable four-loop diagrams where three-loop
master integrals enter in products with one-loop integrals having poles in $\ep$, Laporta
calculated numerically~\cite{Laporta:2001rc} several higher order terms of the
$\ep$-expansion of the three-loop $g-2$ master integrals, with the use of the method of
Refs.~\cite{Laporta:2001dd,Laporta:2000dc}.

We are more optimistic about the possibility to evaluate four-loop $g-2$ master integrals
analytically. To make a step in this direction we have performed an analytical calculation
corresponding to the numerical calculation of Ref.~\cite{Laporta:2001rc}. More precisely, we
have evaluated analytically higher terms of the $\ep$-expansion of the three-loop $g-2$
master integrals up to transcendentality weight seven which is intrinsic to a four-loop
evaluation of the electron $g-2$. So, the highest terms we present involve
$\zeta_7,\zeta_{5}\pi^2, \mbox{Li}_7(1/2), \ln^7 2, \zeta_{-5,1,1}$, etc. In our calculation we
use the DRA method,
obtain results for the master integrals with a very
high precision ($\sim 350-500$ digits) and then apply the so-called PSLQ algorithm
\cite{PSLQ} to arrive at analytical values. There are fifty four independent transcendental
constants in the four loop basis so that this high precision is mandatory. (Typically, one needs
the accuracy of at least seven digits per constant. Observe that this high accuracy is not
provided by the method of Ref.~\cite{Laporta:2001dd,Laporta:2000dc}.)

The three-loop $g-2$ master integrals were analytically evaluated in
Refs.~\cite{Laporta:1996mq,Laporta:1997zy,Melnikov:2000qh,Melnikov:2000zc}. In fact, in
\cite{Melnikov:2000qh,Melnikov:2000zc}, another problem was considered: the evaluation of
three-loop renormalization constants, for which master integrals are almost the same, up to
one additional master integral. This three-loop evaluation was typically performed up to
transcendentality weight five, with some exceptions where some master integrals were
expanded in $\ep$ up to weight four.

In the next two sections we present results for the master integrals for the three-loop
form factors and for the $g-2$ factor, respectively.

\section{Master integrals for quark and gluon \\ form factors}

\begin{figure}
  \includegraphics[width=16cm]{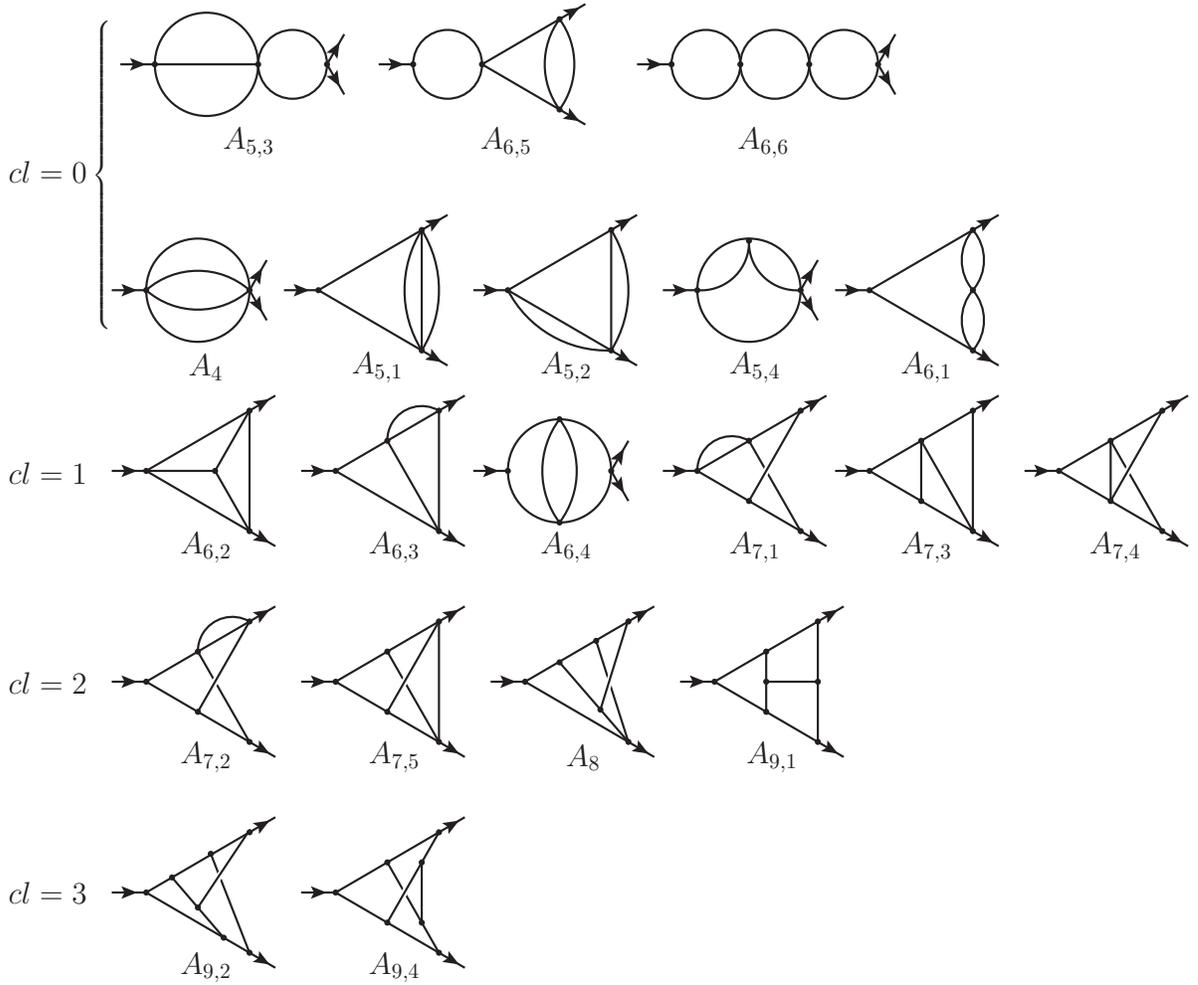}\\
  \caption{Master integrals for the three-loop form factors. Internal lines denote massless
  propagators $1/(k^2+i 0)$.}\label{fig:FFMIs}
\end{figure}

The master integrals for the three-loop form factors are shown in Fig.~\ref{fig:FFMIs}.
Two external momenta are on the light cone, $p_1^2=p_2^2=0$.
In Ref.~\cite{Lee:2010ug} the DRA method 
was applied to the evaluation
of these master integrals. The corresponding results for arbitrary $d$ are obtained in the form
of multiple well-convergent series. Here we present our analytical results on the higher terms
of the $\ep$ expansion which were obtained by calculating these series with a high precision
and then applying the PSLQ algorithm \cite{PSLQ}.

\begin{figure}
  \includegraphics[width=14cm]{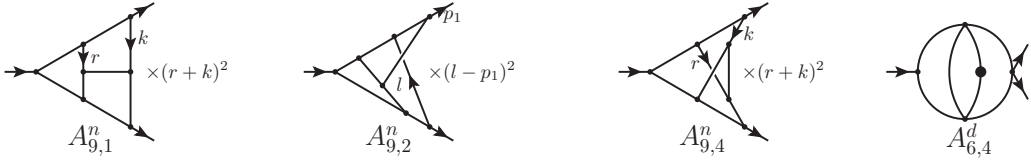}\\
  \caption{Master integrals with a homogeneous transcendentality weight to replace the corresponding
  integrals without numerator.}\label{fig:FFMIsn}
\end{figure}
The high orders of $\epsilon$-expansion allowed us to guess, for all master integrals, except
$A_{6,4}$ and the three most complicated ones ($A_{9,1}$, $A_{9,2}$, and $A_{9,4}$), the
factors which makes the expansion homogeneous in the transcendentality weight. Taking into
account the fact that these four master integrals can be replaced by the integrals with
numerators \cite{3lff4} (or denominator squared for $A_{6,4}$), see Fig. \ref{fig:FFMIsn},
which are also homogeneous in the transcendentality weight, we have a complete basis of
master integrals with homogeneous transcendentality weight. Let us present the results for the
expansion of the integrals in this basis. We arrange the results in the order of increasing
complexity level, the notion introduced in Ref.~\cite{Lee:2010ug}.
The loop integration measure is taken as
\begin{equation}
\frac{d^dk}{i\pi^{d/2}}.
\end{equation}

\noindent\textbf{Zero complexity level}\\
Master integrals with the zero complexity level nullify upon shrinking of any internal line. As we have already argued in Ref.~\cite{Lee:2010ug}, they are always expressed in terms of the $\Gamma$-functions. We present their expansion here only for convenience of the reader.
\begin{eqnarray}
A_{5,3}(4-2\epsilon)&=&-\frac{\Gamma (1-\epsilon )^5 \Gamma (\epsilon ) \Gamma (2 \epsilon -1)}{\Gamma (3-3 \epsilon ) \Gamma (2-2 \epsilon )}
\nonumber\\
&=&
\frac{e^{-3 \gamma  \epsilon }}{(1-3 \epsilon ) (1-2 \epsilon )^2 (1-3 \epsilon/2)}
\Biggl\{
   \frac{1}{4 \epsilon ^2}-\frac{\pi ^2}{16}-\frac{13 \epsilon  \zeta _3}{4}-\frac{17 \pi ^4 \epsilon ^2}{384}
   \nonumber\\&&
   -\epsilon ^3 \biggl(\frac{303 \zeta _5}{20}
   -\frac{13 \pi ^2 \zeta _3}{16}\biggr)
   +\epsilon ^4 \biggl(\frac{169 \zeta _3^2}{8}-\frac{9439 \pi ^6}{483840}\biggr)
   +\epsilon ^5 \biggl(\frac{221 \pi ^4 \zeta _3}{384}
   \nonumber\\&&
   +\frac{303 \pi ^2 \zeta _5}{80}-\frac{2439 \zeta _7}{28}\biggr)
   +\epsilon ^6 \biggl(-\frac{169}{32} \pi ^2 \zeta _3^2+\frac{3939 \zeta _5 \zeta _3}{20}-\frac{76397 \pi ^8}{7741440}\biggr)
   \nonumber\\&&
   +O\left(\epsilon ^7\right)
\Biggr\}\;,
\end{eqnarray}
\begin{eqnarray}
A_{6,5}(4-2\epsilon)&=&\frac{\Gamma (1-2 \epsilon )^2 \Gamma (1-\epsilon )^4 \Gamma (\epsilon )^2 \Gamma (2 \epsilon )}{\Gamma (2-3 \epsilon ) \Gamma (2-2 \epsilon )^2}
\nonumber\\
&=&
\frac{e^{-3 \gamma  \epsilon }}{(1-3 \epsilon ) (1-2 \epsilon )^2}
\Biggl\{
   \frac{1}{2 \epsilon ^3}+\frac{\pi ^2}{24 \epsilon }-\frac{11 \zeta _3}{2}-\frac{77 \pi ^4 \epsilon }{960}
   -\epsilon ^2 \biggl(\frac{11 \pi ^2 \zeta _3}{24}
   \nonumber\\&&
   +\frac{273 \zeta _5}{10}\biggr)
   -\epsilon ^3 \biggl(\frac{5233 \pi ^6}{80640}-\frac{121 \zeta _3^2}{4}\biggr)
   -\epsilon ^4 \biggl(-\frac{847}{960} \pi ^4 \zeta _3+\frac{91 \pi ^2 \zeta _5}{40}
   \nonumber\\&&
   +\frac{2313 \zeta _7}{14}\biggr)
   +\epsilon ^5 \biggl(\frac{121}{48} \pi ^2 \zeta _3^2+\frac{3003 \zeta _5 \zeta _3}{10}-\frac{2326579 \pi ^8}{58060800}\biggr)
   \nonumber\\&&
   +O\left(\epsilon ^6\right)
\Biggr\}
\;,
\end{eqnarray}
\begin{eqnarray}
A_{6,6}(4-2\epsilon)&=&\frac{\Gamma (1-\epsilon )^6 \Gamma (\epsilon )^3}{\Gamma (2-2 \epsilon )^3}
\nonumber\\
&=&
\frac{e^{-3 \gamma  \epsilon }}{(1-2 \epsilon )^3}
\Biggl\{
   \frac{1}{\epsilon ^3}-\frac{\pi ^2}{4 \epsilon }-7 \zeta _3-\frac{37 \pi ^4 \epsilon }{480}
   +\epsilon ^2 \biggl(\frac{7 \pi ^2 \zeta _3}{4}-\frac{93 \zeta _5}{5}\biggr)
   \nonumber\\&&
   +\epsilon ^3 \biggl(\frac{49 \zeta _3^2}{2}-\frac{943 \pi ^6}{120960}\biggr)
   +\epsilon ^4 \biggl(\frac{259 \pi ^4 \zeta _3}{480}+\frac{93 \pi ^2 \zeta _5}{20}-\frac{381 \zeta _7}{7}\biggr)
   \nonumber\\&&
   +\epsilon ^5 \biggl(-\frac{49}{8} \pi ^2 \zeta _3^2+\frac{651 \zeta _5 \zeta _3}{5}+\frac{6527 \pi ^8}{9676800}\biggr)+O\left(\epsilon ^6\right)
\Biggr\}
\;,
\end{eqnarray}
\begin{eqnarray}
A_{4}(4-2\epsilon)&=&\frac{\Gamma (1-\epsilon )^4 \Gamma (3 \epsilon -2)}{\Gamma (4-4 \epsilon )}
\nonumber\\
&=&
\frac{e^{-3 \gamma  \epsilon }}{(1-4 \epsilon ) (3-4 \epsilon ) (1-3 \epsilon ) (2-3 \epsilon ) (1-2 \epsilon )}
\Biggl\{
   \frac{1}{6 \epsilon }-\frac{\pi ^2 \epsilon }{24}-\frac{29 \epsilon ^2 \zeta _3}{6}
   \nonumber\\&&-\frac{71 \pi ^4 \epsilon ^3}{960}
   -\epsilon ^4 \biggl(\frac{421 \zeta _5}{10}-\frac{29 \pi ^2 \zeta _3}{24}\biggr)
   +\epsilon ^5 \biggl(\frac{841 \zeta _3^2}{12}-\frac{11539 \pi ^6}{145152}\biggr)
   \nonumber\\&&
   -\epsilon ^6 \biggl(-\frac{2059}{960} \pi ^4 \zeta _3-\frac{421 \pi ^2 \zeta _5}{40}+\frac{6189 \zeta _7}{14}\biggr)
   +\epsilon ^7 \biggl(-\frac{841}{48} \pi ^2 \zeta _3^2+\frac{12209 \zeta _5 \zeta _3}{10}
   \nonumber\\&&
   -\frac{737687 \pi ^8}{8294400}\biggr)+O\left(\epsilon ^8\right)
\Biggr\}
\;,
\end{eqnarray}
\begin{eqnarray}
A_{5,1}(4-2\epsilon)&=&-\frac{\Gamma (2-3 \epsilon )^2 \Gamma (1-\epsilon )^3 \Gamma (2 \epsilon -1) \Gamma (3 \epsilon -1)}{\Gamma (3-4 \epsilon ) \Gamma (3-3 \epsilon )}
\nonumber\\
&=&
\frac{e^{-3 \gamma  \epsilon }}{(1-4 \epsilon ) (2-3 \epsilon ) (1-2 \epsilon )^2}
\Biggl\{
   -\frac{1}{12 \epsilon ^2}-\frac{\pi ^2}{16}+\frac{23 \epsilon  \zeta _3}{12}-\frac{7 \pi ^4 \epsilon ^2}{1152}
   \nonumber\\&&
   +\epsilon ^3 \biggl(\frac{23 \pi ^2 \zeta _3}{16}+\frac{351 \zeta _5}{20}\biggr)
   +\epsilon ^4 \biggl(\frac{65243 \pi ^6}{1451520}-\frac{529 \zeta _3^2}{24}\biggr)
   +\epsilon ^5 \biggl(\frac{161 \pi ^4 \zeta _3}{1152}
   \nonumber\\&&
   +\frac{1053 \pi ^2 \zeta _5}{80}+\frac{5503 \zeta _7}{28}\biggr)
   +\epsilon ^6 \biggl(-\frac{529}{32} \pi ^2 \zeta _3^2-\frac{8073 \zeta _5 \zeta _3}{20}+\frac{75527 \pi ^8}{860160}\biggr)
   \nonumber\\&&
   +O\left(\epsilon ^7\right)
\Biggr\}
\;,
\end{eqnarray}
\begin{eqnarray}
A_{5,2}(4-2\epsilon)&=&-\frac{\Gamma (2-3 \epsilon ) \Gamma (1-2 \epsilon ) \Gamma (1-\epsilon )^4 \Gamma (\epsilon ) \Gamma (3 \epsilon -1)}{\Gamma (3-4 \epsilon ) \Gamma
   (2-2 \epsilon )^2}
\nonumber\\
&=&
\frac{e^{-3 \gamma  \epsilon }}{(1-4 \epsilon ) (1-2 \epsilon )^3}
\Biggl\{
   \frac{1}{6 \epsilon ^2}+\frac{\pi ^2}{24}-\frac{23 \epsilon  \zeta _3}{6}-\frac{25 \pi ^4 \epsilon ^2}{576}
   -\epsilon ^3 \biggl(\frac{23 \pi ^2 \zeta _3}{24}
   \nonumber\\&&
   +\frac{351 \zeta _5}{10}\biggr)
   -\epsilon ^4 \biggl(\frac{66041 \pi ^6}{725760}-\frac{529 \zeta _3^2}{12}\biggr)
   -\epsilon ^5 \biggl(-\frac{575}{576} \pi ^4 \zeta _3+\frac{351 \pi ^2 \zeta _5}{40}
   \nonumber\\&&
   +\frac{5503 \zeta _7}{14}\biggr)
   -\epsilon ^6 \biggl(-\frac{529}{48} \pi ^2 \zeta _3^2-\frac{8073 \zeta _5 \zeta _3}{10}+\frac{1513373 \pi ^8}{11612160}\biggr)
   \nonumber\\&&
   +O\left(\epsilon ^7\right)
\Biggr\}
\;,
\end{eqnarray}
\begin{eqnarray}
A_{5,4}(4-2\epsilon)&=&-\frac{\Gamma (2-3 \epsilon ) \Gamma (1-\epsilon )^5 \Gamma (\epsilon )^2 \Gamma (3 \epsilon -1)}{\Gamma (3-4 \epsilon ) \Gamma (2-2 \epsilon )^2
   \Gamma (2 \epsilon )}
\nonumber\\
&=&
\frac{e^{-3 \gamma  \epsilon }}{(1-4 \epsilon ) (1-2 \epsilon )^3}
\Biggl\{
   \frac{1}{3 \epsilon ^2}-\frac{\pi ^2}{12}-\frac{23 \epsilon  \zeta _3}{3}-\frac{11 \pi ^4 \epsilon ^2}{96}
   -\epsilon ^3 \biggl(\frac{351 \zeta _5}{5}
   \nonumber\\&&
   -\frac{23 \pi ^2 \zeta _3}{12}\biggr)
   -\epsilon ^4 \biggl(\frac{49199 \pi ^6}{362880}-\frac{529 \zeta _3^2}{6}\biggr)
   -\epsilon ^5 \biggl(-\frac{253}{96} \pi ^4 \zeta _3-\frac{351 \pi ^2 \zeta _5}{20}
   \nonumber\\&&
   +\frac{5503 \zeta _7}{7}\biggr)
   +\epsilon ^6 \biggl(-\frac{529}{24} \pi ^2 \zeta _3^2+\frac{8073 \zeta _5 \zeta _3}{5}-\frac{1006669 \pi ^8}{5806080}\biggr)
   \nonumber\\&&
   +O\left(\epsilon ^7\right)
\Biggr\}
\;,
\end{eqnarray}
\begin{eqnarray}
A_{6,1}(4-2\epsilon)&=&\frac{\Gamma (1-3 \epsilon )^2 \Gamma (1-\epsilon )^4 \Gamma (\epsilon )^2 \Gamma (3 \epsilon )}{\Gamma (2-4 \epsilon ) \Gamma (2-2 \epsilon )^2}
\nonumber\\
&=&
\frac{e^{-3 \gamma  \epsilon }}{(1-4 \epsilon ) (1-2 \epsilon )^2}
\Biggl\{
   \frac{1}{3 \epsilon ^3}+\frac{\pi ^2}{4 \epsilon }-\frac{17 \zeta _3}{3}+\frac{83 \pi ^4 \epsilon }{1440}
   -\epsilon ^2 \biggl(\frac{17 \pi ^2 \zeta _3}{4}
   \nonumber\\&&
   +\frac{281 \zeta _5}{5}\biggr)
   -\epsilon ^3 \biggl(\frac{44651 \pi ^6}{362880}-\frac{289 \zeta _3^2}{6}\biggr)
   -\epsilon ^4 \biggl(\frac{1411 \pi ^4 \zeta _3}{1440}+\frac{843 \pi ^2 \zeta _5}{20}
   \nonumber\\&&
   +\frac{4817 \zeta _7}{7}\biggr)
   -\epsilon ^5 \biggl(-\frac{289}{8} \pi ^2 \zeta _3^2-\frac{4777 \zeta _5 \zeta _3}{5}+\frac{960457 \pi ^8}{3225600}\biggr)
   \nonumber\\&&
   +O\left(\epsilon ^6\right)
\Biggr\}
\;,
\end{eqnarray}

\noindent\textbf{Non-zero complexity level}\\
Integrals with complexity level equal to $n>0$ are expressed in arbitrary dimension $d$ in
terms of an $n$-fold series. We present these integrals in the same order as in
Fig. \ref{fig:FFMIs}. We also present results for $A_{6,4}^{d}$, $A_{9,1}^{n}$, $A_{9,2}^{n}$,
and $A_{9,3}^{n}$ which have the uniform transcendentality and can be used as master integrals
instead of $A_{6,4}$, $A_{9,1}$, $A_{9,2}$, and $A_{9,3}$, respectively.
\begin{eqnarray}
A_{6,2}(4-2\epsilon)&=&\frac{e^{-3 \gamma  \epsilon }}{(1-5 \epsilon ) (1-4 \epsilon )}
\Biggl\{
   \frac{2 \zeta _3}{\epsilon }+\frac{7 \pi ^4}{180}
   -\epsilon  \biggl(\frac{7 \pi ^2 \zeta _3}{6}-10 \zeta _5\biggr)
   \nonumber\\&&
   -\epsilon ^2 \biggl(78 \zeta _3^2+\frac{473 \pi ^6}{15120}\biggr)
   -\epsilon ^3 \biggl(\frac{415 \pi ^4 \zeta _3}{144}+\frac{5 \pi ^2 \zeta _5}{2}+\frac{445 \zeta _7}{2}\biggr)
   \nonumber\\&&
   -\epsilon ^4 \biggl(-\frac{992}{3} \zeta _{-6,-2}-\frac{253}{6} \pi ^2 \zeta _3^2+\frac{2576 \zeta _5 \zeta _3}{5}+\frac{425689 \pi ^8}{1814400}\biggr)
   \nonumber\\&&
   +O\left(\epsilon ^5\right)
\Biggr\}\;,
\end{eqnarray}
\begin{eqnarray}
A_{6,3}(4-2\epsilon)&=&\frac{e^{-3 \gamma  \epsilon }}{(1-4 \epsilon ) (1-3 \epsilon ) (1-2 \epsilon )}
\Biggl\{
   \frac{1}{6 \epsilon ^3}+\frac{\pi ^2}{8 \epsilon }-\frac{35 \zeta _3}{6}-\frac{77 \pi ^4 \epsilon }{2880}
   \nonumber\\&&
   -\epsilon ^2 \biggl(\frac{49 \pi ^2 \zeta _3}{24}+\frac{651 \zeta _5}{10}\biggr)
   +\epsilon ^3 \biggl(\frac{1141 \zeta _3^2}{12}-\frac{93451 \pi ^6}{725760}\biggr)
   \nonumber\\&&
   -\epsilon ^4 \biggl(\frac{713 \pi ^2 \zeta _5}{40}-\frac{511}{320} \pi ^4 \zeta _3+\frac{9017 \zeta _7}{14}\biggr)
   +\epsilon ^5 \biggl(\frac{623}{48} \pi ^2 \zeta _3^2
   \nonumber\\&&
   -\frac{544}{9} \zeta _{-6,-2}+\frac{11195 \zeta _5 \zeta _3}{6}-\frac{2022493 \pi ^8}{11612160}\biggr)
   +O\left(\epsilon ^6\right)
\Biggr\}\;,
\end{eqnarray}
\begin{eqnarray}
A_{6,4}^{d}(4-2\epsilon)&=&
\frac{6 e^{-3 \gamma  \epsilon }}{1-2 \epsilon }
\Biggl\{
   \frac{\zeta _3}{\epsilon }+\frac{\pi ^4}{60}
   +\epsilon  \biggl(-\frac{1}{4} \pi ^2 \zeta _3+17 \zeta _5\biggr)
   +\epsilon ^2 \biggl(\frac{577 \pi ^6}{15120}-20 \zeta _3^2\biggr)
   \nonumber\\&&
   +\epsilon ^3 \biggl(-\frac{301}{480} \pi ^4 \zeta _3-\frac{17 \pi ^2 \zeta _5}{4}+\frac{471 \zeta _7}{2}\biggr)
   +\epsilon ^4 \biggl(\frac{25579 \pi ^8}{604800}+5 \pi ^2 \zeta _3^2
   \nonumber\\&&
   -\frac{1838 \zeta _3 \zeta _5}{5}+48 \zeta _{-6,-2}\biggr)+O\left(\epsilon ^5\right)
\Biggr\}
\;,
\end{eqnarray}
\begin{eqnarray}
A_{6,4}(4-2\epsilon)&=&
e^{-3 \gamma  \epsilon }
\Biggl\{
   \frac{1}{3 \epsilon ^3}+\frac{7}{3 \epsilon ^2}
   +\epsilon ^{-1}\left(\frac{31}{3}-\frac{\pi ^2}{12}\right)+\biggl(\frac{103}{3}-\frac{7 \pi ^2}{12}+\frac{7 \zeta _3}{3}\biggr)
   \nonumber\\&&
   +\epsilon  \biggl(\frac{235}{3}-\frac{31 \pi ^2}{12}+\frac{49 \zeta _3}{3}+\frac{5 \pi ^4}{96}\biggr)
   +\epsilon ^2 \biggl(\frac{19}{3}-\frac{103 \pi ^2}{12}+\frac{289 \zeta _3}{3}
   \nonumber\\&&
   +\frac{35 \pi ^4}{96}-\frac{7 \pi ^2 \zeta _3}{12}+\frac{599 \zeta _5}{5}\biggr)
   +\epsilon ^3 \biggl(-\frac{3953}{3}-\frac{235 \pi ^2}{12}+\frac{1729 \zeta _3}{3}
   \nonumber\\&&
   +\frac{967 \pi ^4}{480}-\frac{49 \pi ^2 \zeta _3}{12}+\frac{4193 \zeta _5}{5}+\frac{108481 \pi ^6}{362880}-\frac{599 \zeta _3^2}{6}\biggr)+\epsilon ^4 \biggl(-\frac{31889}{3}
   \nonumber\\&&
   -\frac{19 \pi ^2}{12}+\frac{10213 \zeta _3}{3}+\frac{5263 \pi ^4}{480}-\frac{289 \pi ^2 \zeta _3}{12}+\frac{20609 \zeta _5}{5}+\frac{108481 \pi ^6}{51840}
   \nonumber\\&&
   -\frac{4193 \zeta _3^2}{6}-\frac{1553 \pi ^4 \zeta _3}{480}-\frac{599 \pi ^2 \zeta _5}{20}+\frac{13593 \zeta _7}{7}\biggr)
   +\epsilon ^5 \biggl(-\frac{188141}{3}
   \nonumber\\&&
   +\frac{3953 \pi ^2}{12}+\frac{57445 \zeta _3}{3}+\frac{28723 \pi ^4}{480}-\frac{1729 \pi ^2 \zeta _3}{12}+\frac{90257 \zeta _5}{5}
   \nonumber\\&&
   +\frac{3695263 \pi ^6}{362880}-\frac{21449 \zeta _3^2}{6}-\frac{10871 \pi ^4 \zeta _3}{480}-\frac{4193 \pi ^2 \zeta _5}{20}+13593 \zeta _7
   \nonumber\\&&+\frac{1913939 \pi ^8}{5806080}+\frac{599}{24} \pi ^2 \zeta _3^2-\frac{9847 \zeta _3 \zeta _5}{5}+576 \zeta _{-6,-2}\biggr)+O\left(\epsilon ^6\right)
\Biggr\}
\;,
\end{eqnarray}
\begin{eqnarray}
A_{7,1}(4-2\epsilon)&=&
\frac{e^{-3 \gamma  \epsilon }}{1-2 \epsilon }
\Biggl\{
   -\frac{1}{4 \epsilon ^5}+\frac{11 \pi ^2}{48 \epsilon ^3}+\frac{41 \zeta _3}{4 \epsilon ^2}+\frac{227 \pi ^4}{1152 \epsilon }+\biggl(\frac{1763 \zeta _5}{20}-\frac{355 \pi ^2 \zeta _3}{48}\biggr)
   \nonumber\\&&
   -\epsilon  \biggl(\frac{1649 \zeta _3^2}{8}-\frac{32759 \pi ^6}{483840}\biggr)
   -\epsilon ^2 \biggl(\frac{43847 \pi ^4 \zeta _3}{5760}+\frac{4971 \pi ^2 \zeta _5}{80}-\frac{16397 \zeta _7}{28}\biggr)
   \nonumber\\&&
   -\epsilon ^3 \biggl(-352 \zeta _{-6,-2}-\frac{12251}{96} \pi ^2 \zeta _3^2+\frac{60603 \zeta _5 \zeta _3}{20}+\frac{23355197 \pi ^8}{116121600}\biggr)
   \nonumber\\&&
   +O\left(\epsilon ^4\right)
\Biggr\}\;,
\end{eqnarray}
\begin{eqnarray}
A_{7,2}(4-2\epsilon)&=&\frac{e^{-3 \gamma  \epsilon }}{1-2 \epsilon }
\Biggl\{
   -\frac{\pi ^2}{12 \epsilon ^3}-\frac{2 \zeta _3}{\epsilon ^2}-\frac{17 \pi ^4}{180 \epsilon }+\biggl(\frac{9 \pi ^2 \zeta _3}{4}-15 \zeta _5\biggr)
   \nonumber\\&&
   +\epsilon  \biggl(75 \zeta _3^2-\frac{1199 \pi ^6}{362880}\biggr)
   +\epsilon ^2 \biggl(\frac{2959 \pi ^4 \zeta _3}{720}+\frac{262 \pi ^2 \zeta _5}{15}+\frac{1383 \zeta _7}{8}\biggr)
   \nonumber\\&&
   +\epsilon ^3 \biggl(-\frac{2912}{9} \zeta _{-6,-2}-\frac{883}{24} \pi ^2 \zeta _3^2+\frac{12493 \zeta _5 \zeta _3}{15}+\frac{21377 \pi ^8}{75600}\biggr)
   \nonumber\\&&
   +O\left(\epsilon ^4\right)
\Biggr\}\;,
\end{eqnarray}
\begin{eqnarray}
A_{7,3}(4-2\epsilon)&=&e^{-3 \gamma  \epsilon }
\Biggl\{
   \epsilon ^{-1}\biggl(\frac{\pi ^2 \zeta _3}{6}+10 \zeta _5\biggr)+\biggl(\frac{31 \zeta _3^2}{2}+\frac{119 \pi ^6}{2160}\biggr)
   \nonumber\\&&
   +\epsilon  \biggl(\frac{61 \pi ^4 \zeta _3}{360}+\frac{23 \pi ^2 \zeta _5}{3}+\frac{1279 \zeta _7}{4}\biggr)
   +\epsilon ^2 \biggl(\frac{656}{3} \zeta _{-6,-2}
   \nonumber\\&&
   -\frac{149}{24} \pi ^2 \zeta _3^2+775 \zeta _5 \zeta _3+\frac{48707 \pi ^8}{604800}\biggr)+O\left(\epsilon ^3\right)
\Biggr\}\;,
\end{eqnarray}
\begin{eqnarray}
A_{7,4}(4-2\epsilon)&=&\frac{e^{-3 \gamma  \epsilon }}{1-6 \epsilon }
\Biggl\{
   -\frac{6 \zeta _3}{\epsilon ^2}-\frac{11 \pi ^4}{90 \epsilon }-\biggl(46 \zeta _5-\frac{7 \pi ^2 \zeta _3}{2}\biggr)
   \nonumber\\&&
   +\epsilon  \biggl(288 \zeta _3^2+\frac{109 \pi ^6}{1080}\biggr)
   +\epsilon ^2 \biggl(\frac{8381 \pi ^4 \zeta _3}{720}+\frac{11 \pi ^2 \zeta _5}{2}+1309 \zeta _7\biggr)
   \nonumber\\&&
   +\epsilon ^3 \biggl(-\frac{7712}{3} \zeta _{-6,-2}-160 \pi ^2 \zeta _3^2+\frac{3668 \zeta _5 \zeta _3}{5}+\frac{72323 \pi ^8}{43200}\biggr)
   \nonumber\\&&
   +O\left(\epsilon ^4\right)
\Biggr\}\;,
\end{eqnarray}
\begin{eqnarray}
A_{7,5}(4-2\epsilon)&=&
\frac{e^{-3 \gamma  \epsilon }}{1-6 \epsilon }
\Biggl\{
   -\biggl(2 \pi ^2 \zeta _3+10 \zeta _5\biggr)
   -\epsilon  \biggl(18 \zeta _3^2+\frac{11 \pi ^6}{162}\biggr)
   \nonumber\\&&
   -\epsilon ^2 \biggl(\frac{13 \pi ^4 \zeta _3}{30}-\frac{7 \pi ^2 \zeta _5}{6}+\frac{307 \zeta _7}{2}\biggr)
   +\epsilon ^3 \biggl(\frac{992}{9} \zeta _{-6,-2}
   \nonumber\\&&
   +\frac{185}{2} \pi ^2 \zeta _3^2+\frac{1690 \zeta _5 \zeta _3}{3}-\frac{2431 \pi ^8}{680400}\biggr)+O\left(\epsilon ^4\right)
\Biggr\}\;,
\end{eqnarray}
\begin{eqnarray}
A_{8}(4-2\epsilon)&=&
\frac{e^{-3 \gamma  \epsilon }}{3 \epsilon +1}
\Biggl\{
   -\frac{8 \zeta _3}{3 \epsilon ^2}-\frac{5 \pi ^4}{27 \epsilon }-\biggl(\frac{352 \zeta _5}{3}-\frac{58 \pi ^2 \zeta _3}{9}\biggr)
   \nonumber\\&&
   +\epsilon  \biggl(\frac{340 \zeta _3^2}{3}-\frac{5261 \pi ^6}{34020}\biggr)
   +\epsilon ^2 \biggl(\frac{5431 \pi ^4 \zeta _3}{540}+\frac{256 \pi ^2 \zeta _5}{9}-\frac{2443 \zeta _7}{3}\biggr)
   \nonumber\\&&
   +\epsilon ^3 \biggl(-\frac{9728}{9} \zeta _{-6,-2}-\frac{1775}{9} \pi ^2 \zeta _3^2+\frac{9968 \zeta _5 \zeta _3}{5}+\frac{1625251 \pi ^8}{4082400}\biggr)
   \nonumber\\&&
   +O\left(\epsilon ^4\right)
\Biggr\}\;,
\end{eqnarray}
\begin{eqnarray}
A_{9,1}^{n}(4-2\epsilon)&=&e^{-3\gamma\epsilon}
\Biggl\{
    \frac{1}{36 \epsilon ^6}+\frac{7 \pi ^2}{144 \epsilon ^4}+\frac{55 \zeta_{3}}{36 \epsilon ^3}+\frac{5329 \pi ^4}{51840 \epsilon ^2}
    +\epsilon^{-1}\biggl(\frac{1171 \pi ^2 \zeta_{3}}{432}+\frac{1199 \zeta_{5}}{60}\biggr)
    \nonumber\\&&
    +\biggl(\frac{353 \zeta_{3}^2}{8}+\frac{2606843 \pi ^6}{13063680}\biggr)
    +\epsilon\biggl(\frac{61831 \pi ^4 \zeta_{3}}{51840}+\frac{46657 \pi ^2 \zeta_{5}}{2160}+\frac{180625 \zeta_{7}}{252}\biggr)
    \nonumber\\&&
    +\epsilon ^2\biggl(-\frac{23680}{81}\zeta_{-6,-2}+\frac{85381 \zeta_{3} \zeta_{5}}{108}-\frac{80579 \pi ^2 \zeta_{3}^2}{864}+\frac{1527522487 \pi ^8}{3135283200}\biggr)
    \nonumber\\&&
    +O\left(\epsilon^3\right)
\Biggr\}\;,
\end{eqnarray}
\begin{eqnarray}
A_{9,1}(4-2\epsilon)&=&e^{-3 \gamma  \epsilon }
\Biggl\{
   \frac{1}{18 \epsilon ^5}
   -\frac{1}{2 \epsilon ^4}
   +\epsilon ^{-3}\biggl(\frac{53}{18}+\frac{29 \pi ^2}{216}\biggr)
   -\epsilon ^{-2}\biggl(\frac{29}{2}+\frac{149 \pi ^2}{216}-\frac{35 \zeta _3}{18}\biggr)
   \nonumber\\&&
   +\epsilon^{-1}\biggl(\frac{129}{2}+\frac{139 \pi ^2}{72}-\frac{307 \zeta _3}{18}+\frac{5473 \pi ^4}{25920}\biggr)
   -\biggl(\frac{537}{2}+\frac{19 \pi ^2}{8}-\frac{1103 \zeta _3}{18}
   \nonumber\\&&
   +\frac{3125 \pi ^4}{5184}-\frac{871 \pi ^2 \zeta _3}{216}-\frac{793 \zeta _5}{10}\biggr)
   +\epsilon  \biggl(\frac{2133}{2}-\frac{97 \pi ^2}{8}-\frac{287 \zeta _3}{2}+\frac{4717 \pi ^4}{2880}
   \nonumber\\&&
   +\frac{2969 \pi ^2 \zeta _3}{216}-\frac{8251 \zeta _5}{30}+\frac{76801 \pi ^6}{186624}+\frac{5521 \zeta _3^2}{36}\biggr)
   -\epsilon ^2 \biggl(\frac{8181}{2}-\frac{969 \pi ^2}{8}
   \nonumber\\&&
   -\frac{195 \zeta _3}{2}+\frac{1333 \pi ^4}{320}+\frac{5887 \pi ^2 \zeta _3}{72}-\frac{22487 \zeta _5}{30}+\frac{4286603 \pi ^6}{6531840}-\frac{799 \zeta _3^2}{4}
   \nonumber\\&&
   -\frac{138403 \pi ^4 \zeta _3}{25920}+\frac{11987 \pi ^2 \zeta _5}{1080}-\frac{228799 \zeta _7}{126}\biggr)
   +\epsilon ^3 \biggl(\frac{30537}{2}-\frac{5589 \pi ^2}{8}
   \nonumber\\&&
   +\frac{2685 \zeta _3}{2}+\frac{9163 \pi ^4}{960}+\frac{2047 \pi ^2 \zeta _3}{8}-\frac{14613 \zeta _5}{10}+\frac{9379 \pi ^6}{16128}-\frac{73819 \zeta _3^2}{36}
   \nonumber\\&&
   +\frac{125549 \pi ^4 \zeta _3}{25920}+\frac{295667 \pi ^2 \zeta _5}{1080}-\frac{542683 \zeta _7}{126}+\frac{325463531 \pi ^8}{313528320}
   \nonumber\\&&
   -\frac{70259}{432} \pi ^2 \zeta _3^2+\frac{38845 \zeta _3 \zeta _5}{54}-\frac{78272}{81} \zeta _{-6,-2}\biggr)+O\left(\epsilon ^4\right)
\Biggr\}\;,
\end{eqnarray}
\begin{eqnarray}
A_{9,2}^{n}(4-2\epsilon)&=&e^{-3\gamma\epsilon}
\Biggl\{
    \frac{2}{9 \epsilon ^6}-\frac{17 \pi ^2}{54 \epsilon ^4}-\frac{31 \zeta_{3}}{3 \epsilon ^3}-\frac{119 \pi ^4}{432 \epsilon ^2}
    +\biggl(\frac{341 \pi ^2 \zeta_{3}}{36}-\frac{2507 \zeta_{5}}{15}\biggr)\epsilon^{-1}
    \nonumber\\&&
    +\biggl(29 \zeta_{3}^2-\frac{195551 \pi ^6}{544320}\biggr)
    +\biggl(-\frac{5963 \pi ^4 \zeta_{3}}{4320}+\frac{8183 \pi ^2 \zeta_{5}}{60}-\frac{43329 \zeta_{7}}{14}\biggr) \epsilon
    \nonumber\\&&
    +\epsilon ^2 \biggl(\frac{20752}{9}\zeta_{-6,-2}+\frac{101288 \zeta_{3} \zeta_{5}}{15}-\frac{6419 \pi ^2 \zeta_{3}^2}{36}-\frac{24178127 \pi ^8}{14515200}\biggr)
    \nonumber\\&&
    +O\left(\epsilon ^3\right)
\Biggr\}\;,
\end{eqnarray}
\begin{eqnarray}
A_{9,2}(4-2\epsilon)&=&e^{-3 \gamma  \epsilon }
\Biggl\{
   -\frac{2}{9 \epsilon ^6}-\frac{5}{6 \epsilon ^5}
   +\epsilon ^{-4}\biggl(\frac{20}{9}+\frac{17 \pi ^2}{54}\biggr)
   +\epsilon ^{-3}\biggl(-\frac{50}{9}+\frac{181 \pi ^2}{216}
   \nonumber\\&&
   +\frac{31 \zeta _3}{3}\biggr)+\epsilon ^{-2}\biggl(\frac{110}{9}-\frac{17 \pi ^2}{9}+\frac{347 \zeta _3}{18}+\frac{119 \pi ^4}{432}\biggr)
   +\epsilon ^{-1}\biggl(-\frac{170}{9}
   \nonumber\\&&
   +\frac{19 \pi ^2}{6}-\frac{514 \zeta _3}{9}+\frac{163 \pi ^4}{960}-\frac{341 \pi ^2 \zeta _3}{36}+\frac{2507 \zeta _5}{15}\biggr)
   +\biggl(-\frac{130}{9}+\frac{\pi ^2}{2}
   \nonumber\\&&
   +\frac{1516 \zeta _3}{9}-\frac{943 \pi ^4}{1080}-\frac{737 \pi ^2 \zeta_3}{24}+\frac{2783 \zeta _5}{6}+\frac{195551 \pi ^6}{544320}-29 \zeta _3^2\biggr)
   \nonumber\\&&
   +\epsilon
   \biggl(\frac{2950}{9}-\frac{83 \pi ^2}{2}-\frac{4444 \zeta _3}{9}+\frac{8801 \pi ^4}{2160}
   +\frac{1357 \pi ^2 \zeta _3}{18}-\frac{2830 \zeta _5}{3}
   \nonumber\\&&
   +\frac{2416889 \pi ^6}{2177280}+\frac{19169 \zeta _3^2}{36}+\frac{5963 \pi ^4 \zeta _3}{4320}
   -\frac{8183 \pi ^2 \zeta _5}{60}+\frac{43329 \zeta _7}{14}\biggr)
   \nonumber\\&&
   +\epsilon ^2
   \biggl(-\frac{19090}{9}+\frac{569 \pi ^2}{2}
   +\frac{12916 \zeta _3}{9}-\frac{7795 \pi ^4}{432}-\frac{1433 \pi ^2 \zeta _3}{9}+\frac{3112 \zeta _5}{3}
   \nonumber\\&&
   -\frac{64733 \pi ^6}{54432}-\frac{1214 \zeta _3^2}{9}+\frac{58517 \pi ^4 \zeta _3}{1728}
   -\frac{146521 \pi ^2 \zeta _5}{360}+\frac{580805 \zeta _7}{42}
   \nonumber\\&&
   +\frac{24178127 \pi ^8}{14515200}+\frac{6419}{36} \pi ^2 \zeta _3^2
   -\frac{101288 \zeta _3 \zeta _5}{15}-\frac{20752}{9} \zeta _{-6,-2}\biggr)
   \nonumber\\&&
   +O\left(\epsilon ^3\right)
\Biggr\}\;,
\end{eqnarray}
\begin{eqnarray}
A_{9,4}^{n}(4-2\epsilon)&=&e^{-3\gamma\epsilon}
\Biggl\{
    -\frac{1}{9 \epsilon ^6}+\frac{43 \pi ^2}{108 \epsilon ^4}+\frac{109 \zeta_{3}}{9 \epsilon ^3}-\frac{481 \pi ^4}{12960 \epsilon ^2}
    +\epsilon^{-1}\biggl(\frac{3463 \zeta_{5}}{45}-\frac{2975 \pi ^2 \zeta_{3}}{108}\biggr)
    \nonumber\\&&
    +\biggl(-\frac{3115 \zeta_{3}^2}{6}-\frac{247613 \pi ^6}{466560}\biggr)
    +\epsilon\biggl(-\frac{38903\pi ^4 \zeta_{3}}{2592}-\frac{113629 \pi ^2 \zeta_{5}}{540}+\frac{8564 \zeta_{7}}{63}\biggr)
    \nonumber\\&&
    +\epsilon ^2 \biggl(\frac{76288}{81} \zeta_{-6,-2}-\frac{730841 \zeta_{3} \zeta_{5}}{135}+\frac{152299 \pi ^2 \zeta_{3}^2}{216}-\frac{30535087 \pi^8}{31352832}\biggr)
    \nonumber\\&&
    +O\left(\epsilon ^3\right)
\Biggr\}\;,
\end{eqnarray}
\begin{eqnarray}
A_{9,4}(4-2\epsilon)&=&e^{-3 \gamma  \epsilon }
\Biggl\{
   -\frac{1}{9 \epsilon ^6}-\frac{8}{9 \epsilon ^5}
   +\epsilon ^{-4}\biggl(1+\frac{43 \pi ^2}{108}\biggr)
   +\epsilon ^{-3}\biggl(\frac{14}{9}+\frac{53 \pi ^2}{27}+\frac{109 \zeta _3}{9}\biggr)
   \nonumber\\&&
   +\epsilon ^{-2}\biggl(-17-\frac{311 \pi ^2}{108}+\frac{608 \zeta _3}{9}-\frac{481 \pi ^4}{12960}\biggr)
   -\epsilon ^{-1}\biggl(-84-\frac{11 \pi ^2}{18}
   \nonumber\\&&
   +\frac{949 \zeta _3}{9}-\frac{85 \pi ^4}{108}+\frac{2975 \pi ^2 \zeta _3}{108}-\frac{3463 \zeta _5}{45}\biggr)
   -\biggl(339-\frac{77 \pi ^2}{4}-\frac{434 \zeta _3}{9}
   \nonumber\\&&
   +\frac{2539 \pi ^4}{2592}+\frac{299 \pi ^2 \zeta _3}{3}-\frac{7868 \zeta _5}{15}+\frac{247613 \pi ^6}{466560}+\frac{3115 \zeta _3^2}{6}\biggr)
   \nonumber\\&&
   -\epsilon  \biggl(-1242+112 \pi ^2-589 \zeta _3+\frac{487 \pi ^4}{432}-\frac{19499 \pi ^2 \zeta _3}{108}+\frac{30067 \zeta _5}{45}
   \nonumber\\&&
   +\frac{25567 \pi ^6}{30240}+\frac{18512 \zeta _3^2}{9}+\frac{38903 \pi ^4 \zeta _3}{2592}+\frac{113629 \pi ^2 \zeta _5}{540}-\frac{8564 \zeta _7}{63}\biggr)
   \nonumber\\&&
   -\epsilon ^2 \biggl(4293-\frac{1887 \pi ^2}{4}+3756 \zeta _3-\frac{491 \pi ^4}{32}+\frac{4019 \pi ^2 \zeta _3}{18}+\frac{7874 \zeta _5}{15}
   \nonumber\\&&
   -\frac{9901847 \pi ^6}{3265920}-\frac{26291 \zeta _3^2}{6}+\frac{9037 \pi ^4 \zeta _3}{135}+\frac{35728 \pi ^2 \zeta _5}{45}-\frac{72537 \zeta _7}{14}
   \nonumber\\&&
   +\frac{30535087 \pi ^8}{31352832}-\frac{152299}{216} \pi ^2 \zeta _3^2+\frac{730841 \zeta _3 \zeta _5}{135}-\frac{76288}{81} \zeta _{-6,-2}\biggr)
   \nonumber\\&&
   +O\left(\epsilon ^3\right)
\Biggr\}\;.
\end{eqnarray}
Here  $\gamma$ is the Euler constant, $\zeta_m=\zeta(m)$, and
$\zeta_{m_1,m_2}=\zeta(m_1,m_2)$ are multiple zeta values (see, e.g., Ref.~\cite{BBV})
\begin{equation}\label{MZVdef}
\zeta(m_1,\dots,m_k)=\sum\limits_{i_1=1}^\infty\sum\limits_{i_2=1}^{i_1-1}
\dots\sum\limits_{i_k=1}^{i_{k-1}-1}\prod\limits_{j=1}^k\frac{\mbox{sgn}(m_j)^{i_j}}{i_j^{|m_j|}}\,.
\nn
\end{equation}

\section{Master integrals for the $g-2$ factor}

\begin{figure}
  \includegraphics[width=16cm]{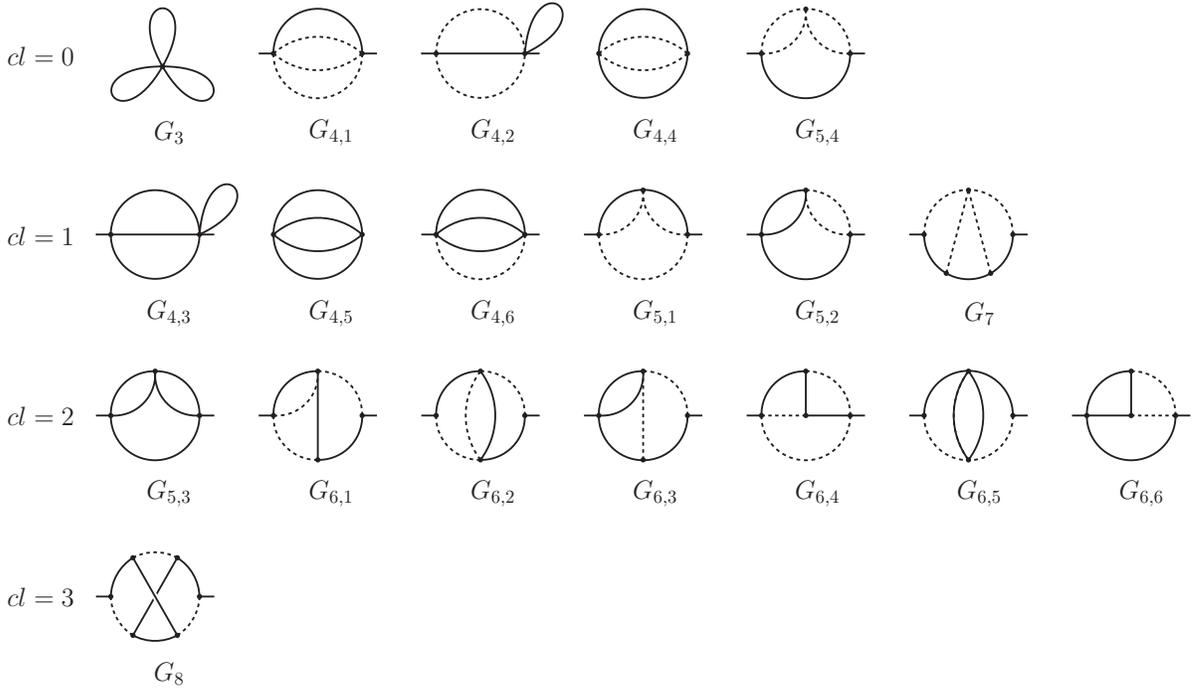}
  \\
  \caption{Master integrals for the three-loop $g-2$ factor. Each line corresponds to a definite complexity level indicated to the left.}
\label{fig:MIs}
\end{figure}

The master integrals for the three-loop $g-2$ factor are shown in Fig.~\ref{fig:MIs}.
The solid internal
lines correspond to massive propagators $1/(-k^2+m^2)$, while the dotted lines correspond to massless
ones, $1/(-k^2)$. The external solid lines denote the incoming and outgoing on-shell momentum $p$,
so that $p^2=m^2$. We set $m=1$ for simplicity. Again, the ordering in Fig.~\ref{fig:MIs} corresponds to increasing complexity level.

\begin{figure}
  \includegraphics[width=7cm]{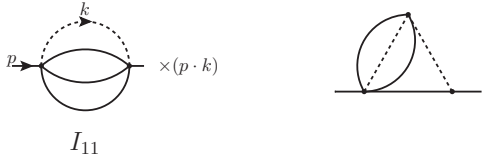}
  \\
  \caption{The integral $I_{11}$ (left) and the auxiliary diagram used for its reduction (right).}
\label{fig:Enveloping}
\end{figure}

It is known \cite{Laporta:1996mq,Laporta:1997zy} that at the three-loop level the  $g - 2$
factor is determined by seventeen master integrals. In fact, in an earlier paper
\cite{Laporta:1996mq} there was one more master integral ($I_{11}$ in the notation of
\cite{Laporta:1996mq}) having the same set of denominators as $G_{4,6}$ --- see Fig.
\ref{fig:Enveloping}.
 In a later paper \cite{Laporta:1997zy} it was recognized that this additional integral can be
expressed via $G_{4,4}$ and $G_{3}$, at least, up to a high power of $\ep$. The DRA method
is based on the consideration of relations exact in $d$, so that we needed an exact relation
between $I_{11}$, $G_{4,4}$, and $G_{3}$. For this purpose, we have performed an IBP
reduction of the integrals with denominators determined by the diagram depicted in Fig.
\ref{fig:Enveloping}. This reduction gave us the expected relation:
\begin{equation}
I_{11}(d)=\frac{2 d-5}{2d-4}G_{4,4}(d)-\frac{1}{4}G_{3}(d)\,.
\end{equation}
Since we are oriented at a future four-loop calculation, we consider two more integrals, $G_{5,4}$
and $G_7$, which also appear in the factorizable four-loop master integrals.
The application of the DRA method allowed us to express all master integrals in $d$ dimensions
in terms of one-,two-, and threefold series with maximal nested depth equal to the complexity level
of the integral. These representations are available upon request from the authors. Here we
present expansions of the master integrals near $d=4$ obtained by numerical summation of the series followed by the application of the PSLQ algorithm.  Again, for convenience of the reader, we start from listing the results for zero complexity level master integrals expressible via $\Gamma$-function.
\begin{eqnarray}
&&G_{3}(4-2\epsilon)=\Gamma (\epsilon-1)^3
\;,
\end{eqnarray}
\begin{eqnarray}
&&G_{4,1}(4-2\epsilon)=\frac{\Gamma (5-6 \epsilon ) \Gamma (1-\epsilon )^3 \Gamma (2 \epsilon -1) \Gamma (3 \epsilon -2)}{\Gamma (4-4 \epsilon ) \Gamma (3-3 \epsilon )}
\nonumber\\&&
=\frac{(1-6 \epsilon ) \Gamma (1+\epsilon)^3}{(1-4 \epsilon ) (3-4 \epsilon ) (1-3 \epsilon ) (2-3 \epsilon ) (1-2 \epsilon )}
\Biggl\{
   -\frac{1}{\epsilon ^2}-2 \pi ^2-32 \epsilon  \zeta _3-\frac{74 \pi ^4 \epsilon ^2}{15}
   \nonumber\\&&
   -\epsilon ^3 \biggl(64 \pi ^2 \zeta _3+1248 \zeta _5\biggr)
   -\epsilon ^4 \biggl(\frac{4636 \pi ^6}{315}+512 \zeta _3^2\biggr)
   -\epsilon ^5 \biggl(\frac{2368 \pi ^4 \zeta _3}{15}+2496 \pi ^2 \zeta _5
   \nonumber\\&&
   +37008 \zeta _7\biggr)+O\left(\epsilon ^6\right)
\Biggr\}
\;,
\end{eqnarray}
\begin{eqnarray}
&&G_{4,2}(4-2\epsilon)=\frac{\Gamma (3-4 \epsilon ) \Gamma (1-\epsilon )^2 \Gamma (\epsilon -1) \Gamma (\epsilon ) \Gamma (2 \epsilon -1)}{\Gamma (3-3 \epsilon ) \Gamma
   (2-2 \epsilon )}
\nonumber\\&&
=\frac{(1-4 \epsilon ) \Gamma (1+\epsilon)^3}{(1-3 \epsilon ) (2-3 \epsilon ) (1-2 \epsilon ) (1
   -\epsilon )}
\Biggl\{
   \frac{1}{\epsilon ^3}+\frac{2 \pi ^2}{3 \epsilon }+8 \zeta _3+\frac{32 \pi ^4 \epsilon }{45}
   +\epsilon ^2 \biggl(\frac{16 \pi ^2 \zeta _3}{3}
   \nonumber\\&&
   +144 \zeta _5\biggr)
   +\epsilon ^3 \biggl(\frac{916 \pi ^6}{945}+32 \zeta _3^2\biggr)
   +\epsilon ^4 \biggl(\frac{256 \pi ^4 \zeta _3}{45}+96 \pi ^2 \zeta _5+1992 \zeta _7\biggr)
   \nonumber\\&&
   +O\left(\epsilon ^5\right)
\Biggr\}
\;,
\end{eqnarray}
\begin{eqnarray}
&&G_{4,4}(4-2\epsilon)=\frac{2^{1-2 \epsilon } \Gamma (2-\epsilon ) \Gamma (\epsilon -1)^2 \Gamma \left(\epsilon -\frac{1}{2}\right) \Gamma (3 \epsilon -2)}{\Gamma \left(2
   \epsilon -\frac{1}{2}\right)}
\nonumber\\&&
=
\frac{(1-4 \epsilon ) \Gamma (1+\epsilon)^3}{(1-3 \epsilon ) (2-3 \epsilon ) (1-2 \epsilon ) (1
   -\epsilon )}
\Biggl\{
   \frac{2}{3 \epsilon ^3}+\frac{16 \zeta _3}{3}-\frac{4 \pi ^4 \epsilon }{15}+96 \epsilon ^2 \zeta _5
   \nonumber\\&&
   -\epsilon ^3 \biggl(\frac{8 \pi ^6}{21}-\frac{64 \zeta _3^2}{3}\biggr)
   +\epsilon ^4 \biggl(-\frac{32}{15} \pi ^4 \zeta _3+1328 \zeta _7\biggr)+O\left(\epsilon ^5\right)
\Biggr\}
\end{eqnarray}
\begin{eqnarray}
&&G_{5,4}(4-2\epsilon)=\frac{\Gamma (3-6 \epsilon ) \Gamma (1-\epsilon )^4 \Gamma (\epsilon )^2 \Gamma (3 \epsilon -1)}{\Gamma (3-4 \epsilon ) \Gamma (2-2 \epsilon )^2}
\nonumber\\&&
=
\frac{(1-6 \epsilon ) \Gamma (1+\epsilon)^3}{(1-4 \epsilon ) (1-2 \epsilon )^3}
\Biggl\{
   -\frac{1}{3 \epsilon ^3}-\frac{2 \pi ^2}{3 \epsilon }-\frac{38 \zeta _3}{3}-\frac{151 \pi ^4 \epsilon }{90}
   -\epsilon ^2 \biggl(\frac{76 \pi ^2 \zeta _3}{3}+430 \zeta _5\biggr)
   \nonumber\\&&
   -\epsilon ^3 \biggl(\frac{4729 \pi ^6}{945}+\frac{722 \zeta _3^2}{3}\biggr)
   -\epsilon ^4 \biggl(\frac{2869 \pi ^4 \zeta _3}{45}+860 \pi ^2 \zeta _5+12434 \zeta _7\biggr)+O\left(\epsilon ^5\right)
\Biggr\}
\end{eqnarray}
\begin{eqnarray}
&&G_{4,3}(4-2\epsilon)=\Gamma (1+\epsilon)^3
\Biggl\{
   \frac{3}{2 \epsilon ^3}+\frac{23}{4 \epsilon ^2}+\frac{105}{8 \epsilon }+\biggl(\frac{275}{16}+\frac{4 \pi ^2}{3}\biggr)+\epsilon  \biggl(-\frac{567}{32}+10 \pi ^2
   \nonumber\\&&
   \phantom{\frac||}-8 \pi ^2 \ln\!2\,+28 \zeta _3\biggr)
   +\epsilon ^2 \biggl(-\frac{14917}{64}+\frac{145 \pi ^2}{3}-60 \pi ^2 \ln\!2\,+210 \zeta _3-\frac{62 \pi ^4}{45}+16 \pi ^2 \ln^2\!2\,   \nonumber\\&&
   \phantom{\frac||}+8 \ln^4\!2\,+192 a_4\biggr)
   +\epsilon ^3 \biggl(-\frac{144015}{128}+\frac{385 \pi ^2}{2}-290 \pi ^2 \ln\!2\,+1015 \zeta _3-\frac{31 \pi ^4}{3}
   \nonumber\\&&
   +120 \pi ^2 \ln^2\!2\,+60 \ln^4\!2\,+1440 a_4+\frac{124}{15} \pi ^4 \ln\!2\,-32 \pi ^2 \ln^3\!2\,-\frac{48 \ln^5\!2\,}{5}+1152 a_5
   \nonumber\\&&
   -\frac{40 \pi ^2 \zeta _3}{3}-930 \zeta _5\biggr)+\epsilon ^4 \biggl(-\frac{1108525}{256}+\frac{8281 \pi ^2}{12}
   -1155 \pi ^2 \ln\!2\,+\frac{8085 \zeta _3}{2}-\frac{899 \pi ^4}{18}
   \nonumber\\&&
   +580 \pi ^2 \ln^2\!2\,+290 \ln^4\!2\,+6960 a_4+62 \pi ^4 \ln\!2\,-240 \pi ^2 \ln^3\!2\,-72 \ln^5\!2\,+8640 a_5
   \nonumber\\&&
   -100 \pi ^2 \zeta _3-6975 \zeta _5-\frac{562 \pi ^6}{135}-\frac{124}{5} \pi ^4 \ln^2\!2\,+48 \pi ^2 \ln^4\!2\,+\frac{48 \ln^6\!2\,}{5}+6912 a_6
   \nonumber\\&&
   +2880 s_6+80 \pi ^2 \ln\!2\, \zeta _3-1220 \zeta _3^2\biggr) +\epsilon ^5 \biggl(-\frac{7710087}{512}+\frac{18585 \pi ^2}{8}-\frac{8281}{2} \pi ^2 \ln\!2\,   \nonumber\\&&
   +\frac{57967 \zeta _3}{4}-\frac{2387 \pi ^4}{12}
   +2310 \pi ^2 \ln^2\!2\,+1155 \ln^4\!2\,+27720 a_4+\frac{899}{3} \pi ^4 \ln\!2\,   \nonumber\\&&
   -1160 \pi ^2 \ln^3\!2\,-348 \ln^5\!2\,+41760 a_5-\frac{1450 \pi ^2 \zeta _3}{3}-\frac{67425 \zeta _5}{2}-\frac{281 \pi ^6}{9}-186 \pi ^4 \ln^2\!2\,   \nonumber\\&&
   +360 \pi ^2 \ln^4\!2\,+72 \ln^6\!2\,+51840 a_6+21600 s_6+600 \pi ^2 \ln\!2\, \zeta _3-9150 \zeta _3^2
   \nonumber\\&&
   +\frac{784}{45} \pi ^6 \ln\!2\,+\frac{248}{5} \pi ^4 \ln^3\!2\,-\frac{288}{5} \pi ^2 \ln^5\!2\,-\frac{288 \ln^7\!2\,}{35}+41472 a_7-\frac{55680}{7} \ln\!2\, s_6
   \nonumber\\&&
   +\frac{4300 \pi ^4 \zeta _3}{63}-240 \pi ^2 \ln^2\!2\, \zeta _3+\frac{69600}{7} \ln\!2\, \zeta _3^2+\frac{32086 \pi ^2 \zeta _5}{7}+16740 \ln^2\!2\, \zeta _5
   \nonumber\\&&
   -\frac{579651 \zeta _7}{7}+\frac{55680 s_{7a}}{7}-\frac{65280 s_{7b}}{7}\biggr)+O\left(\epsilon ^6\right)
\Biggr\}\;,
\end{eqnarray}
\begin{eqnarray}
&&G_{4,5}(4-2\epsilon)=\Gamma (1+\epsilon)^3
\Biggl\{
   \frac{2}{\epsilon ^3}+\frac{23}{3 \epsilon ^2}+\frac{35}{2 \epsilon }+\frac{275}{12}
   +\epsilon  \biggl(-\frac{189}{8}+\frac{112 \zeta _3}{3}\biggr)
   \nonumber\\&&
   -\epsilon ^2 \biggl(\frac{14917}{48}-280 \zeta _3+\frac{136 \pi ^4}{45}+\frac{32}{3} \pi ^2 \ln^2\!2\,-\frac{32 \ln^4\!2\,}{3}-256 a_4\biggr)
   \nonumber\\&&
   -\epsilon ^3 \biggl(\frac{48005}{32}-\frac{4060 \zeta _3}{3}+\frac{68 \pi ^4}{3}+80 \pi ^2 \ln^2\!2\,-80 \ln^4\!2\,-1920 a_4-\frac{272}{15} \pi ^4 \ln\!2\,   \nonumber\\&&
   -\frac{64}{3} \pi ^2 \ln^3\!2\,+\frac{64 \ln^5\!2\,}{5}-1536 a_5+1240 \zeta _5\biggr)
   -\epsilon ^4 \biggl(\frac{1108525}{192}-5390 \zeta _3+\frac{986 \pi ^4}{9}
   \nonumber\\&&
   +\frac{1160}{3} \pi ^2 \ln^2\!2\,-\frac{1160 \ln^4\!2\,}{3}-9280 a_4-136 \pi ^4 \ln\!2\,-160 \pi ^2 \ln^3\!2\,+96 \ln^5\!2\,   \nonumber\\&&
   -11520 a_5+9300 \zeta _5+\frac{32 \pi ^6}{5}+\frac{272}{5} \pi ^4 \ln^2\!2\,+32 \pi ^2 \ln^4\!2\,-\frac{64 \ln^6\!2\,}{5}-9216 a_6
   \nonumber\\&&
   -3840 s_6+\frac{4880 \zeta _3^2}{3}\biggr)
   -\epsilon ^5 \biggl(\frac{2570029}{128}-\frac{57967 \zeta _3}{3}+\frac{1309 \pi ^4}{3}+1540 \pi ^2 \ln^2\!2\,   \nonumber\\&&
   -1540 \ln^4\!2\,-36960 a_4-\frac{1972}{3} \pi ^4 \ln\!2\,-\frac{2320}{3} \pi ^2 \ln^3\!2\,+464 \ln^5\!2\,-55680 a_5
   \nonumber\\&&
   +44950 \zeta _5+48 \pi ^6+408 \pi ^4 \ln^2\!2\,+240 \pi ^2 \ln^4\!2\,-96 \ln^6\!2\,-69120 a_6-28800 s_6
   \nonumber\\&&
   +12200 \zeta _3^2-\frac{3824}{135} \pi ^6 \ln\!2\,-\frac{544}{5} \pi ^4 \ln^3\!2\,-\frac{192}{5} \pi ^2 \ln^5\!2\,+\frac{384 \ln^7\!2\,}{35}-55296 a_7
   \nonumber\\&&
   +\frac{74240}{7} \ln\!2\, s_6-\frac{720 \pi ^4 \zeta _3}{7}-\frac{92800}{7} \ln\!2\, \zeta _3^2-\frac{130360 \pi ^2 \zeta _5}{21}-22320 \ln^2\!2\, \zeta _5
   \nonumber\\&&
   +\frac{772868 \zeta _7}{7}-\frac{74240 s_{7a}}{7}+\frac{87040 s_{7b}}{7}\biggr)+O\left(\epsilon ^6\right)
\Biggr\}\;,
\end{eqnarray}
\begin{eqnarray}
&&G_{4,6}(4-2\epsilon)=\Gamma (1+\epsilon)^3
\Biggl\{
   \frac{1}{\epsilon ^3}+\frac{7}{2 \epsilon ^2}+\frac{253}{36 \epsilon }+\frac{2501}{216}-\biggl(-\frac{59437}{1296}+\frac{64 \pi ^2}{9}\biggr) \epsilon
   \nonumber\\&&
   -\epsilon ^2 \biggl(-\frac{2831381}{7776}+\frac{2272 \pi ^2}{27}-\frac{256}{3} \pi ^2 \ln\!2\,+\frac{1792 \zeta _3}{9}\biggr)-\epsilon ^3 \biggl(-\frac{117529021}{46656}
   \nonumber\\&&
   +\frac{49840 \pi ^2}{81}-\frac{9088}{9} \pi ^2 \ln\!2\,+\frac{63616 \zeta _3}{27}-\frac{2752 \pi ^4}{135}+\frac{3584}{9} \pi ^2 \ln^2\!2\, +\frac{1024 \ln^4\!2\,}{9}
   \nonumber\\&&
   +\frac{8192 a_4}{3}\biggr)
   -\epsilon ^4 \biggl(-\frac{4081770917}{279936}+\frac{875224 \pi ^2}{243}-\frac{199360}{27} \pi ^2 \ln\!2\,+\frac{1395520 \zeta _3}{81}
   \nonumber\\&&
   -\frac{97696 \pi ^4}{405}+\frac{127232}{27} \pi ^2 \ln^2\!2\,+\frac{36352 \ln^4\!2\,}{27}+\frac{290816 a_4}{9}+\frac{11008}{45} \pi ^4 \ln\!2\,   \nonumber\\&&
   -\frac{14336}{9} \pi ^2 \ln^3\!2\,-\frac{4096 \ln^5\!2\,}{15}+32768 a_5-\frac{1792 \pi ^2 \zeta _3}{9}-\frac{87296 \zeta _5}{3}\biggr)
   \nonumber\\&&
   -\epsilon ^5 \biggl(-\frac{125873914573}{1679616}+\frac{13545868 \pi ^2}{729}-\frac{3500896}{81} \pi ^2 \ln\!2\,+\frac{24506272 \zeta _3}{243}-\frac{428624 \pi ^4}{243}
   \nonumber\\&&
   +\frac{2791040}{81} \pi ^2 \ln^2\!2\,+\frac{797440 \ln^4\!2\,}{81}+\frac{6379520 a_4}{27}+\frac{390784}{135} \pi ^4 \ln\!2\,-\frac{508928}{27} \pi ^2 \ln^3\!2\,   \nonumber\\&&
   -\frac{145408 \ln^5\!2\,}{45}+\frac{1163264 a_5}{3}-\frac{63616 \pi ^2 \zeta _3}{27}-\frac{3099008 \zeta _5}{9}-\frac{93184 \pi ^6}{405}-\frac{22016}{15} \pi ^4 \ln^2\!2\,   \nonumber\\&&
   +\frac{14336}{3} \pi ^2 \ln^4\!2\,+\frac{8192 \ln^6\!2\,}{15}+393216 a_6+180224 s_6+\frac{7168}{3} \pi ^2 \ln\!2\, \zeta _3-\frac{633344 \zeta _3^2}{9}\biggr)
   \nonumber\\&&
   -\epsilon ^6 \biggl(-\frac{3593750577461}{10077696}
   +\frac{193770934 \pi ^2}{2187}-\frac{54183472}{243} \pi ^2 \ln\!2\,+\frac{379284304 \zeta _3}{729}
   \nonumber\\&&
   -\frac{37634632 \pi ^4}{3645}+\frac{49012544}{243} \pi ^2 \ln^2\!2\,+\frac{14003584 \ln^4\!2\,}{243}+\frac{112028672 a_4}{81}+\frac{1714496}{81} \pi ^4 \ln\!2\,   \nonumber\\&&
   -\frac{11164160}{81} \pi ^2 \ln^3\!2\,-\frac{637952 \ln^5\!2\,}{27}+\frac{25518080 a_5}{9}-\frac{1395520 \pi ^2 \zeta _3}{81}-\frac{67981760 \zeta _5}{27}
   \nonumber\\&&
   -\frac{3308032 \pi ^6}{1215}-\frac{781568}{45} \pi ^4 \ln^2\!2\,+\frac{508928}{9} \pi ^2 \ln^4\!2\,+\frac{290816 \ln^6\!2\,}{45}+4653056 a_6
   \nonumber\\&&
   +\frac{6397952 s_6}{3}+\frac{254464}{9} \pi ^2 \ln\!2\, \zeta _3-\frac{22483712 \zeta _3^2}{27}+\frac{719872}{405} \pi ^6 \ln\!2\,+\frac{88064}{15} \pi ^4 \ln^3\!2\,   \nonumber\\&&
   -\frac{57344}{5} \pi ^2 \ln^5\!2\,-\frac{32768 \ln^7\!2\,}{35}+4718592 a_7-\frac{19922944}{21} \ln\!2\, s_6+\frac{1789696 \pi ^4 \zeta _3}{189}
   \nonumber\\&&
   -14336 \pi ^2 \ln^2\!2\, \zeta _3+\frac{24903680}{21} \ln\!2\, \zeta _3^2+\frac{38040320 \pi ^2 \zeta _5}{63}+2095104 \ln^2\!2\, \zeta _5-\frac{72259840 \zeta _7}{7}
   \nonumber\\&&
   +\frac{19922944 s_{7a}}{21}-\frac{25493504 s_{7b}}{21}\biggr)+O\left(\epsilon ^7\right)
\Biggr\}\;,
\end{eqnarray}
\begin{eqnarray}
&&G_{5,1}(4-2\epsilon)=\Gamma (1+\epsilon)^3
\Biggl\{
   -\frac{1}{3 \epsilon ^3}-\frac{5}{3 \epsilon ^2
   }-\epsilon^{-1}\left(4+\frac{2 \pi ^2}{3}\right)-\biggl(-\frac{10}{3}+\frac{7 \pi ^2}{3}+\frac{26 \zeta _3}{3}\biggr)
   \nonumber\\&&
   -\epsilon  \biggl(-\frac{302}{3}+\pi ^2+\frac{94 \zeta _3}{3}+\frac{35 \pi ^4}{18}\biggr)
   -\epsilon ^2 \biggl(-734-\frac{101 \pi ^2}{3}+20 \zeta _3
   \nonumber\\&&
   +\frac{551 \pi ^4}{90}+\frac{76 \pi ^2 \zeta _3}{3}+462 \zeta _5\biggr)
   -\epsilon ^3 \biggl(-\frac{12254}{3}-\frac{775 \pi ^2}{3}-\frac{1232 \zeta _3}{3}
   \nonumber\\&&
   -\frac{28 \pi ^4}{15}+\frac{236 \pi ^2 \zeta _3}{3}+1482 \zeta _5+\frac{2353 \pi ^6}{378}+\frac{482 \zeta _3^2}{3}\biggr)-\epsilon ^4 \biggl(-\frac{60346}{3}
   \nonumber\\&&
   -1383 \pi ^2-\frac{9904 \zeta _3}{3}-\frac{5249 \pi ^4}{45}-32 \pi ^2 \zeta _3-252 \zeta _5+\frac{36031 \pi ^6}{1890}
   \nonumber\\&&
   +\frac{1510 \zeta _3^2}{3}+\frac{3571 \pi ^4 \zeta _3}{45}+894 \pi ^2 \zeta _5+15307 \zeta _7\biggr)+O\left(\epsilon ^5\right)
\Biggr\}\;,
\end{eqnarray}
\begin{eqnarray}
&&G_{5,2}(4-2\epsilon)=
\Gamma (1+\epsilon)^3
\Biggl\{
   -\frac{2}{3 \epsilon ^3}-\frac{10}{3 \epsilon ^2}
   -\epsilon^{-1}\left(\frac{26}{3}+\frac{\pi ^2}{3}\right)-\biggl(2+\frac{11 \pi ^2}{3}+\frac{16 \zeta _3}{3}\biggr)
   \nonumber\\&&
   -\epsilon  \biggl(-\frac{398}{3}+\frac{73 \pi ^2}{3}-16 \pi ^2 \ln\!2\,+\frac{248 \zeta _3}{3}+\frac{13 \pi ^4}{45}\biggr)
   -\epsilon ^2 \biggl(-1038+129 \pi ^2
   \nonumber\\&&
   -160 \pi ^2 \ln\!2\,+\frac{1888 \zeta _3}{3}-\frac{3 \pi ^4}{5}+\frac{128}{3} \pi ^2 \ln^2\!2\,+\frac{64 \ln^4\!2\,}{3}+512 a_4+\frac{8 \pi ^2 \zeta _3}{3}+96 \zeta _5\biggr)
   \nonumber\\&&
   -\epsilon ^3 \biggl(-\frac{17470}{3}+\frac{1817 \pi ^2}{3}-1024 \pi ^2 \ln\!2\,+3600 \zeta _3-\frac{751 \pi ^4}{45}+\frac{1280}{3} \pi ^2 \ln^2\!2\, +\frac{640 \ln^4\!2\,}{3}
   \nonumber\\&&
   +5120 a_4+\frac{736}{45} \pi ^4 \ln\!2\,-\frac{1024}{9} \pi ^2 \ln^3\!2\,-\frac{512 \ln^5\!2\,}{15}+4096 a_5-\frac{8 \pi ^2 \zeta _3}{3}-2496 \zeta _5+\frac{368 \pi ^6}{945}
   \nonumber\\&&
   +\frac{64 \zeta _3^2}{3}\biggr)
   -\epsilon ^4 \biggl(-\frac{85562}{3}+2649 \pi ^2-5376 \pi ^2 \ln\!2\,+\frac{53264 \zeta _3}{3}-\frac{5849 \pi ^4}{45}+\frac{8192}{3} \pi ^2 \ln^2\!2\,   \nonumber\\&&
   +\frac{4096 \ln^4\!2\,}{3}+32768 a_4+\frac{1472}{9} \pi ^4 \ln\!2\,-\frac{10240}{9} \pi ^2 \ln^3\!2\,-\frac{1024 \ln^5\!2\,}{3}+40960 a_5
   \nonumber\\&&
   -\frac{376 \pi ^2 \zeta _3}{3}-28512 \zeta _5-\frac{274 \pi ^6}{15}-\frac{3424}{45} \pi ^4 \ln^2\!2\,+\frac{2144}{9} \pi ^2 \ln^4\!2\,+\frac{2048 \ln^6\!2\,}{45}
   \nonumber\\&&
   +256 \pi ^2 a_4+32768 a_6+12288 s_6+352 \pi ^2 \ln\!2\, \zeta _3-\frac{14680 \zeta _3^2}{3}+\frac{104 \pi ^4 \zeta _3}{45}+48 \pi ^2 \zeta _5
   \nonumber\\&&
   +1328 \zeta _7\biggr)+O\left(\epsilon ^5\right)
\Biggr\}\;,
\end{eqnarray}

\begin{eqnarray}
&&G_{7}(4-2\epsilon)=\Gamma (1+\epsilon)^3
\Biggl\{
   \biggl(2 \pi ^2 \zeta _3-5 \zeta _5\biggr)
   +\epsilon  \biggl(4 \pi ^2 \zeta _3-10 \zeta _5+\frac{16 \pi ^6}{189}+7 \zeta _3^2\biggr)
   \nonumber\\&&
   +\epsilon ^2 \biggl(8 \pi ^2 \zeta _3-20 \zeta _5+\frac{32 \pi ^6}{189}+14 \zeta _3^2-\frac{181 \pi ^4 \zeta _3}{30}+166 \pi ^2 \zeta _5-212 \zeta _7\biggr)
   \nonumber\\&&
   +O\left(\epsilon ^3\right)
\Biggr\}\;,
\end{eqnarray}
\begin{eqnarray}
&&G_{5,3}(4-2\epsilon)=
\Gamma (1+\epsilon)^3
\Biggl\{
   -\frac{1}{\epsilon ^3}-\frac{16}{3 \epsilon ^2}-\frac{16}{\epsilon }-\biggl(20+\frac{8 \pi ^2}{3}-2 \zeta _3\biggr)
   -\epsilon  \biggl(-\frac{364}{3}+28 \pi ^2
   \nonumber\\&&
   +\frac{200 \zeta _3}{3}+\frac{3 \pi ^4}{10}-16 \pi ^2 \ln\!2\,\biggr)
   -\epsilon ^2 \biggl(-1244+188 \pi ^2+776 \zeta _3-\frac{46 \pi ^4}{15}-21 \pi ^2 \zeta _3+126 \zeta _5
   \nonumber\\&&
   -168 \pi ^2 \ln\!2\,+\frac{80}{3} \pi ^2 \ln^2\!2\,+\frac{64 \ln^4\!2\,}{3}+512 a_4\biggr)
   -\epsilon ^3 \biggl(-7572+\frac{3100 \pi ^2}{3}+5360 \zeta _3-\frac{218 \pi ^4}{5}
   \nonumber\\&&
   -\frac{332 \pi ^2 \zeta _3}{3}-1976 \zeta _5-\frac{22 \pi ^6}{35}-332 \zeta _3^2-1128 \pi ^2 \ln\!2\,+\frac{128}{5} \pi ^4 \ln\!2\,+126 \pi ^2 \ln\!2\, \zeta _3
   \nonumber\\&&
   +280 \pi ^2 \ln^2\!2\,-6 \pi ^4 \ln^2\!2\,-\frac{160}{3} \pi ^2 \ln^3\!2\,+224 \ln^4\!2\,+6 \pi ^2 \ln^4\!2\,-\frac{128 \ln^5\!2\,}{5}+5376 a_4
   \nonumber\\&&
   +144 \pi ^2 a_4+3072 a_5\biggr)+O\left(\epsilon ^4\right)
\Biggr\}\;,
\end{eqnarray}
\begin{eqnarray}
&&G_{6,1}(4-2\epsilon)=\Gamma (1+\epsilon)^3
\Biggl\{
   \frac{1}{6 \epsilon ^3}+\frac{3}{2 \epsilon ^2}
   +\epsilon ^{-1}\biggl(\frac{55}{6}-\frac{\pi ^2}{3}\biggr)+\biggl(\frac{95}{2}-2 \pi ^2-\frac{8 \zeta _3}{3}-\frac{\pi ^4}{15}\biggr)
   \nonumber\\&&
   +\epsilon  \biggl(\frac{1351}{6}-\frac{17 \pi ^2}{3}-14 \zeta _3-\frac{47 \pi ^4}{45}+6 \pi ^2 \zeta _3-64 \zeta _5\biggr)
   +\epsilon ^2 \biggl(\frac{2023}{2}+\frac{16 \pi ^2}{3}
   \nonumber\\&&
   -16 \pi ^2 \ln\!2\,+\frac{16 \zeta _3}{3}-\frac{457 \pi ^4}{90}+\frac{26 \pi ^2 \zeta _3}{3}-342 \zeta _5+\frac{1471 \pi ^6}{2835}+\frac{8}{3} \pi ^4 \ln^2\!2\,   \nonumber\\&&
   -\frac{8}{3} \pi ^2 \ln^4\!2\,-64 \pi ^2 a_4-56 \pi ^2 \ln\!2\, \zeta _3+62 \zeta _3^2\biggr)+\epsilon ^3 \biggl(\frac{26335}{6}+187 \pi ^2-224 \pi ^2 \ln\!2\,   \nonumber\\&&
   +598 \zeta _3-\frac{277 \pi ^4}{15}+\frac{224}{3} \pi ^2 \ln^2\!2\,+\frac{64 \ln^4\!2\,}{3}+512 a_4-\frac{58 \pi ^2 \zeta _3}{3}-1082 \zeta _5
   \nonumber\\&&
   -\frac{299 \pi ^6}{378}+8 \pi ^4 \ln^2\!2\,-8 \pi ^2 \ln^4\!2\,-192 \pi ^2 a_4-168 \pi ^2 \ln\!2\, \zeta _3+\frac{400 \zeta _3^2}{3}-\frac{112}{135} \pi ^6 \ln\!2\,   \nonumber\\&&
   -\frac{64}{3} \pi ^4 \ln^3\!2\,+\frac{128}{5} \pi ^2 \ln^5\!2\, +768 \pi ^2 \ln\!2\, a_4+768 \pi ^2 a_5+1024 \ln\!2\, s_6
   \nonumber\\&&
   +\frac{1651 \pi ^4 \zeta _3}{45}+\frac{1232}{3} \pi ^2 \ln^2\!2\, \zeta _3-\frac{224}{3} \ln^4\!2\, \zeta _3-1792 a_4 \zeta _3 -1280 \ln\!2\, \zeta _3^2-\frac{499 \pi ^2 \zeta _5}{3}
   \nonumber\\&&
  -8365 \zeta _7-1024 s_{7a}-1024 s_{7b}\biggr)+O\left(\epsilon ^4\right)
\Biggr\}\;,
\end{eqnarray}
\begin{eqnarray}
&&G_{6,2}(4-2\epsilon)=
\Gamma (1+\epsilon)^3
\Biggl\{
   \frac{1}{3 \epsilon ^3}+\frac{7}{3 \epsilon ^2}+\frac{31}{3 \epsilon }+\biggl(\frac{103}{3}+\frac{\pi ^2}{3}+\frac{2 \zeta _3}{3}-\frac{4 \pi ^4}{45}\biggr)
   +\epsilon  \biggl(\frac{235}{3}
   \nonumber\\&&
   +4 \pi ^2+\frac{20 \zeta _3}{3}-\frac{3 \pi ^4}{10}+\frac{2 \pi ^2 \zeta _3}{3}-2 \zeta _5\biggr)
   +\epsilon ^2 \biggl(\frac{19}{3}+\frac{91 \pi ^2}{3}-16 \pi ^2 \ln\!2\,+\frac{206 \zeta _3}{3}
   \nonumber\\&&
   +\frac{14 \pi ^4}{45}+2 \pi ^2 \zeta _3+6 \zeta _5+\frac{1009 \pi ^6}{1890}+\frac{8}{3} \pi ^4 \ln^2\!2\,-\frac{8}{3} \pi ^2 \ln^4\!2\,-64 \pi ^2 a_4-56 \pi ^2 \ln\!2\, \zeta _3
   \nonumber\\&&
   +32 \zeta _3^2\phantom{\frac||}\biggr)
   +\epsilon ^3 \biggl(-\frac{3953}{3}+186 \pi ^2-224 \pi ^2 \ln\!2\,+\frac{1760 \zeta _3}{3}+\frac{307 \pi ^4}{90}+\frac{224}{3} \pi ^2 \ln^2\!2\,   \nonumber\\&&
   +\frac{64 \ln^4\!2\,}{3}+512 a_4+14 \pi ^2 \zeta _3+222 \zeta _5+\frac{979 \pi ^6}{630}+8 \pi ^4 \ln^2\!2\,-8 \pi ^2 \ln^4\!2\,-192 \pi ^2 a_4
   \nonumber\\&&
   -168 \pi ^2 \ln\!2\, \zeta _3+\frac{296 \zeta _3^2}{3}-\frac{112}{135} \pi ^6 \ln\!2\,-\frac{64}{3} \pi ^4 \ln^3\!2\,+\frac{128}{5} \pi ^2 \ln^5\!2\,+768 \pi ^2 \ln\!2\, a_4
   \nonumber\\&&
   +768 \pi ^2 a_5+1024 \ln\!2\, s_6+\frac{976 \pi ^4 \zeta _3}{45}+\frac{1232}{3} \pi ^2 \ln^2\!2\, \zeta _3-\frac{224}{3} \ln^4\!2\, \zeta _3-1792 a_4 \zeta _3
   \nonumber\\&&
   -1280 \ln\!2\, \zeta _3^2-215 \pi ^2 \zeta _5-6114 \zeta _7-1024 s_{7a}-1024 s_{7b}\biggr)+O\left(\epsilon ^4\right)
\Biggr\}\;,
\end{eqnarray}
\begin{eqnarray}
&&G_{6,3}(4-2\epsilon)=
\Gamma (1+\epsilon)^3
\Biggl\{
   \frac{1}{6 \epsilon ^3}+\frac{3}{2 \epsilon ^2}
   +\epsilon ^{-1}\biggl(\frac{55}{6}-\frac{\pi ^2}{3}\biggr)+\biggl(\frac{95}{2}-\frac{7 \pi ^2}{3}-\frac{14 \zeta _3}{3}-\frac{4 \pi ^4}{45}\biggr)
   \nonumber\\&&
   +\epsilon  \biggl(\frac{1351}{6}-\frac{31 \pi ^2}{3}+4 \pi ^2 \ln\!2\,-42 \zeta _3-\frac{23 \pi ^4}{30}+\frac{25 \pi ^2 \zeta _3}{6}-\frac{49 \zeta _5}{2}\biggr)
   +\epsilon ^2 \biggl(\frac{2023}{2}
   \nonumber\\&&
   -\frac{103 \pi ^2}{3}+32 \pi ^2 \ln\!2\,-\frac{698 \zeta _3}{3}-\frac{148 \pi ^4}{45}-\frac{40}{3} \pi ^2 \ln^2\!2\,-\frac{20 \ln^4\!2\,}{3}-160 a_4+\frac{109 \pi ^2 \zeta _3}{6}
   \nonumber\\&&
   -\frac{413 \zeta _5}{2}+\frac{23 \pi ^6}{54}+\frac{7}{3} \pi ^4 \ln^2\!2\,-\frac{7}{3} \pi ^2 \ln^4\!2\,-56 \pi ^2 a_4-49 \pi ^2 \ln\!2\, \zeta _3+\frac{191 \zeta _3^2}{4}\biggr)
   \nonumber\\&&
   +\epsilon ^3 \biggl(\frac{26335}{6}-\frac{235 \pi ^2}{3}+140 \pi ^2 \ln\!2\,-994 \zeta _3-\frac{371 \pi ^4}{30}-80 \pi ^2 \ln^2\!2\,-56 \ln^4\!2\,   \nonumber\\&&
   -1344 a_4-\frac{182}{45} \pi ^4 \ln\!2\,+\frac{368}{9} \pi ^2 \ln^3\!2\,+\frac{184 \ln^5\!2\,}{15}-1472 a_5+\frac{129 \pi ^2 \zeta _3}{2}+\frac{107 \zeta _5}{2}
   \nonumber\\&&
   +\frac{457 \pi ^6}{270}+\frac{35}{3} \pi ^4 \ln^2\!2\,-\frac{35}{3} \pi ^2 \ln^4\!2\,-280 \pi ^2 a_4-245 \pi ^2 \ln\!2\, \zeta _3+\frac{2641 \zeta _3^2}{12}
   \nonumber\\&&
   -\frac{35}{54} \pi ^6 \ln\!2\,-\frac{164}{9} \pi ^4 \ln^3\!2\,+\frac{328}{15} \pi ^2 \ln^5\!2\,+656 \pi ^2 \ln\!2\, a_4+656 \pi ^2 a_5+800 \ln\!2\, s_6
   \nonumber\\&&
   +\frac{3803 \pi ^4 \zeta _3}{180}+\frac{1036}{3} \pi ^2 \ln^2\!2\, \zeta _3-\frac{175}{3} \ln^4\!2\, \zeta _3-1400 a_4 \zeta _3-1000 \ln\!2\, \zeta _3^2-\frac{2057 \pi ^2 \zeta _5}{12}
   \nonumber\\&&
   -\frac{44413 \zeta _7}{8}-800 s_{7a}-800 s_{7b}\biggr)+O\left(\epsilon ^4\right)
\Biggr\}\;,
\end{eqnarray}
\begin{eqnarray}
&&G_{6,4}(4-2\epsilon)=
\Gamma (1+\epsilon)^3
\Biggl\{
   \frac{2 \zeta _3}{\epsilon }-\biggl(-\frac{\pi ^2}{3}-2 \zeta _3+\frac{7 \pi ^4}{90}\biggr)
   +\epsilon  \biggl(\frac{14 \pi ^2}{3}-12 \zeta _3
   \nonumber\\&&
   -\frac{41 \pi ^4}{90}-\frac{2 \pi ^2 \zeta _3}{3}+44 \zeta _5\biggr)
   -\epsilon ^2 \biggl(-\frac{119 \pi ^2}{3}+16 \pi ^2 \ln\!2\,+38 \zeta _3+\frac{14 \pi ^4}{45}+\frac{26 \pi ^2 \zeta _3}{3}
   \nonumber\\&&
   +42 \zeta _5-\frac{1447 \pi ^6}{2835}-\frac{8}{3} \pi ^4 \ln^2\!2\,+\frac{8}{3} \pi ^2 \ln^4\!2\,+64 \pi ^2 a_4+56 \pi ^2 \ln\!2\, \zeta _3-54 \zeta _3^2\biggr)
   \nonumber\\&&
   +\epsilon ^3 \biggl(\frac{796 \pi ^2}{3}-224 \pi ^2 \ln\!2\,+256 \zeta _3+\frac{51 \pi ^4}{10}+\frac{224}{3} \pi ^2 \ln^2\!2\,+\frac{64 \ln^4\!2\, }{3}+512 a_4
   \nonumber\\&&
   -26 \pi ^2 \zeta _3-574 \zeta _5+\frac{31 \pi ^6}{630}+8 \pi ^4 \ln^2\!2\,-8 \pi ^2 \ln^4\!2\,-192 \pi ^2 a_4-168 \pi ^2 \ln\!2\, \zeta _3
   \nonumber\\&&
   +104 \zeta _3^2-\frac{112}{135} \pi ^6 \ln\!2\,-\frac{64}{3} \pi ^4 \ln^3\!2\,+\frac{128}{5} \pi ^2 \ln^5\!2\,+768 \pi ^2 \ln\!2\, a_4+768 \pi ^2 a_5
   \nonumber\\&&
   +1024 \ln\!2\, s_6+\frac{931 \pi ^4 \zeta _3}{45}+\frac{1232}{3} \pi ^2 \ln^2\!2\, \zeta _3-\frac{224}{3} \ln^4\!2\, \zeta _3-1792 a_4 \zeta _3
   \nonumber\\&&
   -1280 \ln\!2\, \zeta _3^2-217 \pi ^2 \zeta _5-5655 \zeta _7-1024 s_{7a}-1024 s_{7b}\phantom{\frac||}\biggr)+O\left(\epsilon ^4\right)
\Biggr\}\;,
\end{eqnarray}
\begin{eqnarray}
&&G_{6,5}(4-2\epsilon)=
\Gamma (1+\epsilon)^3
\Biggl\{
   \frac{1}{3 \epsilon ^3}+\frac{7}{3 \epsilon ^2}+\frac{31}{3 \epsilon }+\biggl(\frac{103}{3}-\frac{4 \zeta _3}{3}-\frac{2 \pi ^4}{15}\biggr)
   +\epsilon  \biggl(\frac{235}{3}
   \nonumber\\&&
   +\frac{8 \pi ^2}{3}+\frac{32 \zeta _3}{3}-\frac{3 \pi ^4}{5}+\frac{28 \pi ^2 \zeta _3}{3}-78 \zeta _5\biggr)
   +\epsilon ^2 \biggl(\frac{19}{3}+\frac{112 \pi ^2}{3}-32 \pi ^2 \ln\!2\,+\frac{692 \zeta _3}{3}
   \nonumber\\&&
   -\frac{164 \pi ^4}{45}-\frac{16}{3} \pi ^2 \ln^2\!2\,+\frac{16 \ln^4\!2\,}{3}+128 a_4+\frac{140 \pi ^2 \zeta _3}{3}-414 \zeta _5+\frac{928 \pi ^6}{945}
   \nonumber\\&&
   +\frac{16}{3} \pi ^4 \ln^2\!2\,-\frac{16}{3} \pi ^2 \ln^4\!2\,-128 \pi ^2 a_4-112 \pi ^2 \ln\!2\, \zeta _3+169 \zeta _3^2\biggr)+\epsilon ^3 \biggl(-\frac{3953}{3}+\frac{952 \pi ^2}{3}
   \nonumber\\&&
   -448 \pi ^2 \ln\!2\,+\frac{6392 \zeta _3}{3}-\frac{269 \pi ^4}{9}+96 \pi ^2 \ln^2\!2\,+96 \ln^4\!2\,+2304 a_4+\frac{136}{15} \pi ^4 \ln\!2\,   \nonumber\\&&
   +\frac{32}{3} \pi ^2 \ln^3\!2\,-\frac{32 \ln^5\!2\,}{5}+768 a_5+\frac{532 \pi ^2 \zeta _3}{3}-2246 \zeta _5+\frac{946 \pi ^6}{189}+\frac{80}{3} \pi ^4 \ln^2\!2\,   \nonumber\\&&
   -\frac{80}{3} \pi ^2 \ln^4\!2\,-640 \pi ^2 a_4-560 \pi ^2 \ln\!2\, \zeta _3+\frac{2519 \zeta _3^2}{3}-\frac{266}{135} \pi ^6 \ln\!2\,-\frac{128}{3} \pi ^4 \ln^3\!2\,   \nonumber\\&&
   +\frac{256}{5} \pi ^2 \ln^5\!2\,+1536 \pi ^2 \ln\!2\, a_4+1536 \pi ^2 a_5+2432 \ln\!2\, s_6+\frac{367 \pi ^4 \zeta _3}{5}+\frac{2548}{3} \pi ^2 \ln^2\!2\, \zeta _3
   \nonumber\\&&
   -\frac{532}{3} \ln^4\!2\, \zeta _3-4256 a_4 \zeta _3-3040 \ln\!2\, \zeta _3^2-132 \pi ^2 \zeta _5-\frac{35591 \zeta _7}{2}-2432 s_{7a}-2432 s_{7b}\biggr)
   \nonumber\\&&
   +O\left(\epsilon ^4\right)
\Biggr\}\;,
\end{eqnarray}
\begin{eqnarray}
&&G_{6,6}(4-2\epsilon)=
\Gamma (1+\epsilon)^3
\Biggl\{
   \frac{2 \zeta _3}{\epsilon }-\biggl(\frac{\pi ^2}{3}-10 \zeta _3+\frac{13 \pi ^4}{90}\biggr)
   +\epsilon  \biggl(-\frac{4 \pi ^2}{3}+4 \pi ^2 \ln\!2\,   \nonumber\\&&
   +24 \zeta _3-\frac{13 \pi ^4}{18}+\frac{49 \pi ^2 \zeta _3}{6}-\frac{85 \zeta _5}{2}\biggr)
   -\epsilon ^2 \biggl(-7 \pi ^2-16 \pi ^2 \ln\!2\,-42 \zeta _3+\frac{217 \pi ^4}{90}
   \nonumber\\&&
   +\frac{40}{3} \pi ^2 \ln^2\!2\,+\frac{20 \ln^4\!2\,}{3}+160 a_4-\frac{245 \pi ^2 \zeta _3}{6}+\frac{425 \zeta _5}{2}-\frac{1751 \pi ^6}{1890}-5 \pi ^4 \ln^2\!2\,   \nonumber\\&&
   +5 \pi ^2 \ln^4\!2\,+120 \pi ^2 a_4+105 \pi ^2 \ln\!2\, \zeta _3-\frac{495 \zeta _3^2}{4}\biggr)
   +\epsilon ^3 \biggl(\frac{394 \pi ^2}{3}-84 \pi ^2 \ln\!2\,   \nonumber\\&&
   +268 \zeta _3-\frac{223 \pi ^4}{18}-\frac{16}{3} \pi ^2 \ln^2\!2\,-\frac{104 \ln^4\!2\,}{3}-832 a_4-\frac{182}{45} \pi ^4 \ln\!2\,+\frac{368}{9} \pi ^2 \ln^3\!2\,   \nonumber\\&&
   +\frac{184 \ln^5\!2\,}{15}-1472 a_5+\frac{313 \pi ^2 \zeta _3}{2}+\frac{431 \zeta _5}{2}+\frac{1751 \pi ^6}{378}+25 \pi ^4 \ln^2\!2\,-25 \pi ^2 \ln^4\!2\,   \nonumber\\&&
   -600 \pi ^2 a_4-525 \pi ^2 \ln\!2\, \zeta _3+\frac{2475 \zeta _3^2}{4}-\frac{133}{90} \pi ^6 \ln\!2\,-\frac{356}{9} \pi ^4 \ln^3\!2\, +\frac{712}{15} \pi ^2 \ln^5\!2\,   \nonumber\\&&
   +1424 \pi ^2 \ln\!2\, a_4+1424 \pi ^2 a_5+1824 \ln\!2\, s_6+\frac{9259 \pi ^4 \zeta _3}{180}+756 \pi ^2 \ln^2\!2\, \zeta _3-133 \ln^4\!2\, \zeta _3
   \nonumber\\&&
   -3192 a_4 \zeta _3-2280 \ln\!2\, \zeta _3^2-\frac{3841 \pi ^2 \zeta _5}{12}-\frac{105133 \zeta _7}{8}-1824 s_{7a}-1824 s_{7b}\biggr)
   \nonumber\\&&
   +O\left(\epsilon ^4\right)
\Biggr\}\;,
\end{eqnarray}
\begin{eqnarray}
&&G_{8}(4-2\epsilon)=
\Gamma (1+\epsilon)^3
\Biggl\{
   \biggl(-\frac{\pi ^4}{6}+4 \pi ^2 \ln^2\!2\,\biggr)
   +\epsilon  \biggl(\frac{2 \pi ^4}{3}-16 \pi ^2 \ln^2\!2\,
   +\frac{166}{45} \pi ^4 \ln\!2\,
   \nonumber\\&&
   -\frac{208}{9} \pi ^2 \ln^3\!2\,-\frac{32 \ln^5\!2\,}{15}+256 a_5
   +\frac{17 \pi ^2 \zeta _3}{6}-291 \zeta _5\biggr)
   +\epsilon ^2 \biggl(-\frac{14 \pi ^4}{3}+112 \pi ^2 \ln^2\!2\,
   \nonumber\\&&
   -\frac{664}{45} \pi ^4 \ln\!2\,+\frac{832}{9} \pi ^2 \ln^3\!2\,
   +\frac{128 \ln^5\!2\,}{15}-1024 a_5-\frac{34 \pi ^2 \zeta _3}{3}+1164 \zeta _5
   \nonumber\\&&
   -\frac{21743 \pi ^6}{11340}-\frac{197}{9} \pi ^4 \ln^2\!2\,
   +\frac{713}{9} \pi ^2 \ln^4\!2\,+\frac{64 \ln^6\!2\,}{9}-104 \pi ^2 a_4
   +5120 a_6+2688 s_6
   \nonumber\\&&
   -51 \pi ^2 \ln\!2\, \zeta _3-953 \zeta _3^2\biggr)
   +\epsilon ^3 \biggl(\frac{80 \pi ^4}{3}-640 \pi ^2 \ln^2\!2\,
   +\frac{4648}{45} \pi ^4 \ln\!2\,-\frac{5824}{9} \pi ^2 \ln^3\!2\,
   \nonumber\\&&
   -\frac{896 \ln^5\!2\,}{15}+7168 a_5+\frac{238 \pi ^2 \zeta _3}{3}-8148 \zeta _5
   +\frac{21743 \pi ^6}{2835}+\frac{788}{9} \pi ^4 \ln^2\!2\,
   -\frac{2852}{9} \pi ^2 \ln^4\!2\,
   \nonumber\\&&
   -\frac{256 \ln^6\!2\,}{9}+416 \pi ^2 a_4-20480 a_6-10752 s_6
   +204 \pi ^2 \ln\!2\, \zeta _3+3812 \zeta _3^2+\frac{4868}{189} \pi ^6 \ln\!2\,
   \nonumber\\&&
   +\frac{8492}{135} \pi ^4 \ln^3\!2\,-\frac{7288}{45} \pi ^2 \ln^5\!2\,
   -\frac{4864 \ln^7\!2\,}{315}+1776 \pi ^2 \ln\!2\, a_4+1520 \pi ^2 a_5+77824 a_7
   \nonumber\\&&
   -\frac{106880}{7} \ln\!2\, s_6+\frac{4003 \pi ^4 \zeta _3}{21}
   +\frac{2167}{3} \pi ^2 \ln^2\!2\, \zeta _3-\frac{316}{3} \ln^4\!2\, \zeta _3
   -2528 a_4 \zeta _3
   \nonumber\\&&
   +\frac{133600}{7} \ln\!2\, \zeta _3^2+\frac{875561 \pi ^2 \zeta _5}{84}
   +37200 \ln^2\!2\, \zeta _5-\frac{1325727 \zeta _7}{7}+\frac{106880 s_{7a}}{7}
   \nonumber\\&&
   -\frac{161920 s_{7b}}{7}\biggr)+O\left(\epsilon ^4\right)
\Biggr\}\;.
\end{eqnarray}
Here
\bea
a_n&=&\mbox{Li}_n\left(1/2\right)\;,
\nn \\
s_6&=&\zeta_{-5,-1}+\zeta_6\;,
\nn \\
s_{7a}&=&  \zeta_{-5,1,1}+\zeta_{-6,1}+\zeta_{-5,2}+\zeta_{-7}\;,
\nn \\
s_{7b}&=&\zeta_7+\zeta_{5,2}+\zeta_{-6,-1}+\zeta_{5,-1,-1}\;.
\nn
\eea

\section{Conclusion}

Using the DRA method \cite{Lee:2009dh} we have performed the analytic evaluation of two
families of the three-loop master integrals in the $\ep$-expansion up to the transcendentality
weight intrinsic to the corresponding four-loop master integrals. This method is naturally
combined with other methods. First, it hardly relies on IBP reduction which is necessary in
order to obtain difference equations with respect to dimension in a closed form. To do this we
used the code based on  Ref.~\cite{Lee:2008tj} and the code called {\tt FIRE} \cite{FIRE}. To
reveal the position and the order of the poles in a basic stripe we used a sector
decomposition \cite{BH,BognerWeinzierl,FIESTA} implemented in the code {\tt FIESTA}
\cite{FIESTA,FIESTA2}. To fix remaining constants in the homogenous solution of dimensional
recurrence relations we applied the method of Mellin--Barnes representation
\cite{MB1,MB2,books2}. When dealing with multiple zeta values we used the code {\tt HPL}
\cite{Maitre:2005uu}. Finally, to arrive at analytical results for coefficients in the
$\ep$-expansion we applied the PSLQ algorithm \cite{PSLQ}.


\vspace*{2mm}

\noindent
 {\em Acknowledgments.}

This work was supported by the Russian Foundation for Basic Research through grant
08--02--01451. The work of R.L. was also supported through Federal special-purpose program "Scientific and scientific-pedagogical personnel of innovative Russia".

\end{document}